\documentclass[a4paper,11pt]{article}
\usepackage{jcappub} 
\usepackage{lineno}
\usepackage{subcaption}
\usepackage{multirow}
\usepackage{threeparttable}
\usepackage{booktabs}

\newcommand{\ms}{~\mathrm{ms}}
\newcommand{\kpc}{~\mathrm{kpc}}
\newcommand{\erg}{~\mathrm{erg}}
\newcommand{\keV}{~\mathrm{keV}}
\newcommand{\MeV}{~\mathrm{MeV}}

\newcommand{\ee}{\mathrm{e}}

\graphicspath{{./Figures/}}

\title{\boldmath Bayesian Inference of Supernova Neutrino Spectra with Multiple Detectors}

 \author[a,b]{Xu-Run Huang,}
 \author[a]{Chuan-Le Sun,}
 \author[a]{Lie-Wen Chen,}
 \author[a]{Jun Gao}
 \affiliation[a]{School of Physics and Astronomy, Shanghai Key Laboratory for Particle Physics and Cosmology, and Key Laboratory for Particle Astrophysics and Cosmology (MOE),
Shanghai Jiao Tong University,\\ Shanghai 200240, China}
 \affiliation[b]{Department of Physics, The Chinese University of Hong Kong,\\ Shatin, N.T., Hong Kong S.A.R., China}

\emailAdd{xr-huang@cuhk.edu.hk}
\emailAdd{chlsun60@sjtu.edu.cn}
\emailAdd{lwchen@sjtu.edu.cn}
\emailAdd{jung49@sjtu.edu.cn}

\abstract{We implement the Bayesian inference to retrieve energy spectra of all neutrinos from a galactic core-collapse supernova (CCSN). To achieve high statistics and full sensitivity to all flavours of neutrinos, we adopt a combination of several reaction channels from different large-scale neutrino observatories, namely inverse beta decay on proton and elastic scattering on electron from Hyper-Kamiokande (Hyper-K), charged current absorption on Argon from Deep Underground Neutrino Experiment (DUNE) and coherent elastic scattering on Lead from RES-NOVA. Assuming no neutrino oscillation or specific oscillation models, we obtain mock data for each channel through Poisson processes with the predictions, for a typical source distance of 10 kpc in our Galaxy, and then evaluate the probability distributions for all spectral parameters of theoretical neutrino spectrum model with Bayes' theorem. Although the results for either the electron-neutrinos or electron-antineutrinos reserve relatively large uncertainties (according to the neutrino mass ordering), a precision of a few percent (i.e., $\pm 1 \% \sim \pm 4 \%$ at a credible interval of $2 \sigma$) is achieved for primary spectral parameters (e.g., mean energy and total emitted energy) of other neutrino species. Moreover, the correlation coefficients between different parameters are computed as well and interesting patterns are found. Especially, the mixing-induced correlations are sensitive to the neutrino mass ordering, which potentially makes it a brand new probe to determine the neutrino mass ordering in the detection of galactic supernova neutrinos. Finally, we discuss limitations and perspectives for further improvement on our results.}

\begin{document}
\maketitle
\flushbottom

\section{Introduction}

The epochal detection of neutrino signals of SN 1987A, deriving from the Large Magellanic Cloud ($\sim 50 \kpc$), revealed the veil of multi-messenger era of astrophysics. 
Although only about two dozen neutrinos from this transient were caught by three lucky detectors, namely Kamiokande II~\cite{Kamiokande-II:1987idp}, Irvine-Michigan-Brookhaven (IMB)~\cite{Bionta:1987qt} and Baksan~\cite{ALEXEYEV1988209}, this detection renders to us the first glimpse into the collapsing core of a dying massive star. 
After that, various analyses, based on such sparse data, confirmed the outline of stellar core collapse and meanwhile imposed constraints on elusive properties of neutrinos~\cite{Sato:1987rd,Burrows:1987zz,Arnett:1987iz,Bahcall:1987ua,Loredo:1988mk,Loredo:2001rx}. 
Three decades after that landmark, extraordinary progresses have been made among the modelling of stellar core collapse~\cite{Bethe:1990mw,Janka:2012wk,Muller:2016izw,OConnor:2018sti,Just:2018djz,Burrows:2020qrp}, neutrino physics~\cite{Xing:2020ijf} and neutrino detection~\cite{Scholberg:2012id,Mosel:2016cwa,Dutta:2019oaj}. 
That is, millions of neutrinos will be detected with unprecedentedly high precision in modern neutrino observatories if the next galactic CCSN exploded at a typical distance of $\sim 10 \kpc$ (approximately the distance between the centre of the Milky Way and our Solar System)~\cite{Mirizzi:2015eza,Horiuchi:2018ofe}. 
Such detection will promise, with no doubt, a much vaster playground for investigating meaningful topics in both CCSN physics and neutrino physics~\cite{Mirizzi:2015eza,Horiuchi:2018ofe,Muller:2019upo} (also other potentially interesting topics~\cite{Huang:2022wqu,Chauhan:2022wgj,Baum:2022wfc}).

Modern hydrodynamic codes are now capable of performing successful simulations of the collapse and explosion of massive stars~\cite{Muller:2016izw,OConnor:2018tuw,Burrows:2019zce,Nagakura:2020qhb}.
They enrich our understanding of the explosion mechanism and characteristics of the related neutrino emission~\cite{Muller:2019upo,Burrows:2020qrp}.
However, a direct confirmation of those models is still missing and thus highly anticipated.
Multiple neutrino detectors are currently in operation and scrutinizing the cosmos, or expected to operate in the future.
Furthermore, some of them can promise unprecedentedly high statistics if the target is not too far, including water-based Cherenkov detectors (Hyper-Kamiokande~\cite{Hyper-Kamiokande:2018ofw}, IceCube~\cite{IceCube:2011cwc}), liquid scintillator detectors (JUNO~\cite{JUNO:2015zny}, THEIA~\cite{Theia:2019non}), liquid argon time projection chambers (DUNE~\cite{DUNE:2020lwj,DUNE:2020ypp,DUNE:2020zfm}), Pb-based cryogenic detectors (RES-NOVA~\cite{Pattavina:2020cqc,RES-NOVA:2021gqp}) and so on.
Although it is too complicated to predict when the next CCSN will occur in the vicinity, a rate of $1.63 \pm 0.46$ CCSN/100 y is obtained for the Milky Way and galaxies in the Local Group~\cite{Rozwadowska:2020nab}. 
So, it could be promising to anticipate at least one galactic CCSN during the missions of those contemporary or next-generation detectors. 

Such a prospect has attracted quite some attentions on how to maximize the scientific return from such detection in the communities of astrophysics and particle physics.
Among them, reconstructing the energy spectrum of neutrinos is significant for physics but demanding for the amount and quality of data. 
Attributing to the relatively strong interaction and low requirement on detector construction, inverse beta decay on proton (IBD-p) has become the most widely-utilised reaction channel in large-scale neutrino detectors~\cite{JUNO:2015zny,Hyper-Kamiokande:2018ofw,Theia:2019non}.
This literally promises a good sensitivity to electron-antineutrinos. 
Elastic scattering on electron and charged current reaction on nuclei (e.g. $^{12}\mathrm{C}$~\cite{JUNO:2015zny}, $^{16}\mathrm{O}$~\cite{Hyper-Kamiokande:2018ofw} and $^{40}\mathrm{Ar}$~\cite{DUNE:2020ypp,DUNE:2020zfm}) offer the approaches to catch electron-neutrinos.
Previous works have shown that a reasonable precision can be achievable in the measurement of supernova $\nu_e$ spectrum~\cite{Laha:2013hva,Nikrant:2017nya}.
Now, the last task is presented as achieving sufficient sensitivity to heavy flavour neutrinos which can only undergo neutral current processes in such low-energy region. 
Therefore, elastic scattering on proton (pES) in scintillator detectors has been naturally proposed as an available access to heavy flavour part of supernova neutrinos~\cite{Dasgupta:2011wg}.
Nevertheless, the RES-NOVA project, recently proposed in ref.~\cite{Pattavina:2020cqc} with the primary mission of detecting supernova neutrinos, promises high statistics via the coherent elastic neutrino-nucleus scattering on Lead.
Note that different species of heavy flavour neutrinos are generally indistinguishable from each other since none of their charged companions would emerge with sufficiently large amount in stellar core collapse~\footnote{Thus, $\nu_x$ is commonly used to denote one species of heavy flavour neutrinos and so do we. Sometimes, $\nu_x$ and $\bar{\nu}_x$ appear simultaneously, then they indicate particles and anti-particles, respectively.}.
However, a synergy of reaction channels is indispensable for extracting flavour-depending information (e.g. the collection of IBD-p, elastic scattering on electron/proton and charged/neutral current reactions on nuclei~\cite{Lu:2016ipr,GalloRosso:2017mdz,GalloRosso:2020qqa,Li:2017dbg,Li:2019qxi,Nagakura:2020bbw}). 

According to methodology, previous efforts can be schematically divided into two categories: statistical approaches and unfolding processes. 
Based on certain templates, statistical analysis extracts signals from noisy data with high efficiency, and thus has been usually adopted~\cite{Laha:2013hva,Lu:2016ipr,Nikrant:2017nya,GalloRosso:2017mdz,GalloRosso:2020qqa}.
In such analyses, the profiles of neutrino fluxes are commonly depicted by the sophisticated Garching formula~\cite{Keil:2002in}, which has been proven to be well compatible with high-resolution simulations~\cite{Tamborra:2012ac}. 
To some extent, this simple fit represents our sophistication on the modelling of stellar core collapse.
However, the heavy dependence on this analytic formula may potentially discard some important features of the real signals. 
Unfolding methods~\cite{Li:2017dbg,Li:2019qxi,Nagakura:2020bbw} are capable of alleviating such drawback, since they do not rely on any analytical formulas.
But the shortages of such methods are even more severe.
Aside from the complexity, the spectral reversion with response matrix belongs to the case of ill-posed problem, which means that small errors or noise can easily lead to artificial patterns in the results~\cite{Nagakura:2020bbw}.
So, the pragmatic strategy is to implement these two processes complementarily in analysis of supernova neutrinos.
They all offer meaningful information, only in different manner.
In this work, we employ the Bayesian statistics to perform such evaluations.  
In the last decades, Bayesian method~\cite{DAgostini:2003bpu} has been proven to be a powerful tool in questions generally handling uncertainty, including gravitational wave astronomy~\cite{Ashton:2018jfp},  relativistic heavy-ion collisions~\cite{Bernhard:2016tnd}, astrophysics and cosmology~\cite{Trotta:2008qt,Loredo:2012jm}, and fields of human activity beyond fundamental physics (e.g. Bayesian networks). 
Especially, it had already been introduced to the analysis of neutrino signals from SN 1987A~\cite{Loredo:2001rx}.

In this paper, we demonstrate the use of Bayes' theorem to evaluate the spectral parameters for all flavours of neutrinos from a galactic CCSN.
At the source, we adopt the time-integrated spectra for each type of neutrinos from a long-term axisymmetric core-collapse simulation which is reported in ref.~\cite{Nikrant:2017nya}.
Then, the simple adiabatic conversion in the CCSN~\cite{Scholberg:2017czd} is applied here to account for the inevitable oscillation effects, including the case of normal mass ordering (NMO) and inverted mass ordering (IMO). 
We also show the results with no oscillation effects.
However, any other neutrino conversion models can also be implemented in principle.
As to the detection, we attempt to simultaneously obtain high statistics and full sensitivities to all types of neutrinos by taking advantage of three large-scale neutrino observatories, namely Hyper-K, DUNE and RES-NOVA. 
It should also be mentioned that the pES channel in JUNO is capable of performing flavour-blind detection with high energy resolution.  
However, it is reported that the reconstructed $\nu_e$ and $\nu_x$ spectra suffer from a substantial systematic bias of energy threshold induced by the pES channel's insensitivity to neutrinos with energy below $20 \MeV$~\cite{Li:2019qxi}. 
Note that the peak is usually located at $\sim 10 \MeV$ in the spectrum of supernova neutrinos.
Instead, the proposed $1 \keV$ threshold for nuclear recoil energy in RES-NOVA offers the flavour-blind sensitivity to neutrinos with energy above $\sim 10 \MeV$~\cite{Pattavina:2020cqc}.
As a demonstration of our method, we make optimistic assumptions and adopt the RES-NOVA setup in its 3rd phase.
Nevertheless, it is worthy to mention that the pES channel in JUNO may provide valuable all-flavour information for other purposes, or contribute useful all-flavour information here if the RES-NOVA detector of this scale was not realised in a realistic measurement.
Detailed configurations of these detectors will be discussed later.
The fast event-rate calculation tool, \textit{SNOwGLoBES}~\footnote{\textit{SNOwGLoBES} provides detector responses to many reaction channels (see e.g. ref.~\cite{Scholberg:2012id} for details) and it is available at \url{https://webhome.phy.duke.edu/~schol/snowglobes/}.}, is employed to compute count rates for channels in Hyper-K and DUNE, while that for RES-NOVA is done with a code developed by our own~\footnote{This code and \textit{SNOwGLoBES} have been integrated in our Bayesian code.}.
In section \ref{sec:SNv&Dec}, we review the detector characteristics and generate the mock data for further analysis. Aside from the detector responses, noise from Poisson processes is also included in the mock data.
In section \ref{sec:BayInf&NumRes}, we demonstrate how the spectral parameters are estimated from the mock data via Bayes' theorem, and numerical results as well. 
The effects from cross section uncertainties are estimated.
Finally, we conclude this study and discuss the limitation in section \ref{sec:Con}. 

\section{Supernova neutrinos in detectors}
\label{sec:SNv&Dec}

Before getting into details of Bayesian analysis, we summarise the features of detectors employed in this work and the characteristics of supernova neutrinos.
Since no experimental data is available up to now, we calculate the number of expected events in each energy bin for each channels, based on the neutrino fluxes from numerical simulation, and then extract the number of events for analysis from a Poisson distribution with the expected count as average value.
How we consider the neutrino oscillation effects is also presented in this section.

\subsection{Detector configurations}
\label{subsec:DetCon} 

The primary reaction channels for the selected detectors, namely IBD-p in Hyper-K, charged current reaction on Argon (vAr(CC)) in DUNE and neutral current scattering on Lead (vPb(NC)) in RES-NOVA, are adopted in this study to provide sensitivities to $\bar{\nu}_e$, $\nu_e$ and $\nu$, sequentially.
We also include the elastic scattering on electron (eES) in Hyper-K, in order to further enhance the sensitivity of this collection to $\nu_e$ and $\nu_x$. 
Note that eES channel have different cross sections to each type of neutrinos, i.e., $\sigma_{\nu_e} > \sigma_{\bar{\nu}_e} > \sigma_{\nu_x}$\footnote{Strictly speaking, $\sigma_{\nu_x}$ is slightly greater than $\sigma_{\bar{\nu}_x}$ (see figure 2 in ref.~\cite{Scholberg:2012id}).}.
It is also interesting to mention that neutral current scattering on Argon in DUNE can potentially offer good sensitivity to $\nu_x$, just not yet fully studied~\cite{DUNE:2020zfm}.

Hyper-K is a next-generation water-based Cherenkov detector which is scheduled to start data-taking in 2027~\cite{Hyper-Kamiokande:2021frf}.
Its primary missions include precision measurements on neutrino oscillations, searches for proton decay and observations on astrophysical neutrinos~\cite{Hyper-Kamiokande:2018ofw}.
In this study, we employ two reaction channels in Hyper-K, namely the IBD-p ($\bar{\nu}_e + p \to e^+ + n$) and eES ($\nu + e^- \to \nu + e^-$).
Electrons and anti-electrons are produced in these scatterings and emit Cherenkov lights along with their motions in ultra-pure water. 
Then, the events can be reconstructed by collecting those Cherenkov photons via photomultiplier tubes (PMT).
Currently, the reconstruction of IBD-p event has been well established. 
Meanwhile, eES event can also get separated from IBD-p signals, to some extent, according to their different angular dependence. 
Furthermore, it is reported that the neutron tagging efficiency can get improved substantially through addition of gadolinium (e.g., an efficiency of $\sim 90 \%$ in a gadolinium-loaded Super-K)~\cite{Laha:2013hva}. 
That is, the tagging efficiency for the two reaction channels is expected to be promising since the possibility of gadolinium loading has already been considered in the design report of Hyper-K \footnote{The project of loading gadolinium into Super-K has already been approved. And this will provide a template for further application in Hyper-K. See ref.~\cite{Hyper-Kamiokande:2018ofw} for more details.}.
Here we just assume a generally full tagging efficiency for the two reactions.
On the other hand, according to the design report, the fully configured Hyper-K detector consists of two tanks, of which each contains 258 kton of ultra-pure water. 
The designed fiducial mass for each tank reaches 187 kton.
Therefore, a 374 kton of total fiducial mass for Hyper-K has been adopted in some of previous works (see, e.g., ref.~\cite{Nikrant:2017nya,GalloRosso:2017mdz}).
However, the realistic fiducial mass for one tank can exceed this designed scale and reach 220 kton in the detection of supernova neutrinos, because of the localization in time and the neglect of low energy radioactive background due to the short-time feature of supernova neutrino signals~\cite{Hyper-Kamiokande:2018ofw}.
We thus consider one tank with a fiducial mass of 220 kton, just following the available scale also adopted in ref.~\cite{Nagakura:2020bbw}.
That is, only half of the capability of Hyper-K is under evaluation in this study.
As to detector response, we adopt the same smearing matrix and post-smearing efficiency as that of Super-K I (or III, IV), which are provided in \textit{SNOwGLoBES}.
Its response corresponds to the assumption of $40 \%$ PMT coverage.

DUNE~\cite{DUNE:2020lwj,DUNE:2020ypp} will consist of four time projection chambers which contains 70 kton liquid argon in total.
The nominal fiducial mass is 40 kton, and we also adopt this value in this study.
However, in principle the available mass may exceed this value when studying supernova neutrinos, just like the case in Hyper-K.
The primary goals for DUNE include precision measurements on neutrino oscillation parameters and searching for new physics.
Among current-operated and future-planed neutrino detectors, DUNE will bring unique sensitivity to $\nu_e$ with energies down to $\sim 5 \MeV$ via the vAr(CC) reaction ($\nu_e + ^{40}\mathrm{Ar} \to e^- +^{40}\mathrm{K}^\ast$).
When such reactions happen, short electron tracks will be created and recorded, potentially along with gamma-rays in the chambers.
DUNE will also have excellent time resolution which assures its capability of precisely depicting the neutrino burst time profile if the source is close enough.
For instance, it is possible to identify the neutrino ``trapping notch", which emerges as a consequence of neutrino trapping in the dense collapsing core and typically has a width of $1-2 \ms$, for closest CCSNe (few kpc)~\cite{DUNE:2020zfm}.
Moreover, in the galactic supernova neutrino detection landscape with DUNE, one of the most interesting topic is that the mass ordering problem in neutrino oscillations can be decisively determined by the detection of neutronization burst which is almost composed of $\nu_e$ when produced~\cite{Scholberg:2017czd}. 
The above works also adopted \textit{SNOwGLoBES} in their studies.
Therefore, it is quite convenient for us since the configurations of DUNE has already been provided as well.

RES-NOVA~\cite{Pattavina:2020cqc,RES-NOVA:2021gqp} is a newly proposed experiment with the primary aim of hunting neutrinos from CCSNe.
It intends to achieve a flavour-blind measurement with low energy threshold, high energy resolution and high statistics to supernova neutrinos, by taking advantage of the large coherent elastic scattering cross sections between MeV neutrinos and Pb nuclei, the ultrahigh radiopurity of archaeological Pb and modern technologies on cryogenic detector.
This innovative project carries the ambition of providing a $5 \sigma$ sensitivity to supernova bursts up to Andromeda.
However, the detailed configuration has not been settled yet.
In this work, we consider a simple realisation of RN-3 in ref.~\cite{Pattavina:2020cqc}, which is constructed with pure Pb crystals and has a detector mass of 465 ton. 
It will have a $1 \keV$ energy threshold and a $0.2 \keV$ resolution for nuclear recoil energy.
This means that RES-NOVA could be sensitive to neutrinos with energies down to $\sim 10 \MeV$.
However, it should be stressed here that we use the observed nuclear recoil energy when handling the data from RES-NOVA, instead of the reconstructed neutrino energy adopted in previous detectors.
When neutrinos arrive at the detector, they can possibly undergo the vPb(NC) processes ($\nu + \rm{Pb} \to \nu + \rm{Pb}$). After that, the target nucleus will gain a recoil energy in the magnitude of a few keV, and then billions of phonons will get created in the absorber and act as information carriers. 
Such experimental strategy can possibly make full use of the entire energies deposited in the detector and lead to a realisation of excellent energy reconstruction.
However, unlike the previous detectors, the configuration of RES-NOVA is currently absent in \textit{SNOwGLoBES}.
We calculate the event rates following our previous works (i.e., ref.~\cite{Huang:2022wqu,Huang:2019ene}).
The averaged neutron skin of Pb nuclei is fixed on the experimental value of $^{208}\rm{Pb}$, namely $R_n - R_p = 0.283 \pm 0.071 ~\rm{fm}$ from PREX-II~\cite{PREX:2021umo}.
Furthermore, in order to properly account for the effect of threshold,  we adopt such an acceptance efficiency function:
\begin{equation}
\label{eq:efficFunc}
    A(x) = \dfrac{a}{1 + \ee^{- k (x - x_0)}} ,
\end{equation}
where the values of parameters are taken as $a = 1, k = 5, x_0 = 1.5$.
Such arrangements assure that the detection efficiency will swiftly rise up to around 100\% from $\sim 0\%$ when nuclear recoil energy goes to 2 keV from 1 keV, and approaches 100\% asymptotically after 2 keV.
At our estimate, the assumption of a full acceptance efficiency will increase the accuracy of $\alpha_{\nu_x}$ by $\sim 16 \%$ in the case of no neutrino oscillations, owing to the higher statistics in the energy range below $\sim 2 \keV$, while the impact is merely invisible on the extraction of other parameters.
In fact, this function derives from the acceptance efficiency of the COHERENT experiment~\cite{COHERENT:2017ipa,COHERENT:2018imc}, and can also produce similar structure as the reconstruction efficiency function of DUNE~\cite{DUNE:2020zfm}, just with different parameters. 
Note that this efficiency represents a conservative estimate and the real one is yet to be determined. 

\subsection{Neutrino spectra and oscillations}

State-of-the-art stellar evolution theory indicates that dying massive stars would undergo violent core collapse at their end, generating an outward-propagating shock-wave to expel their mantles and exploding as spectacular CCSNe which can emerge as luminous as their host galaxy. 
In such explosions, almost $\sim 99 \%$ of the released gravitational potential energy ($\sim 10^{53} \erg$) will be liberated through neutrino emission.
Moreover, the evolutionary histories of the dense core are imprinted in both the temporal structures and energy spectra of neutrino emissions.
Note that the neutrinos can still deliver information out of the collapsing core, even if no electromagnetic signal was emitted due to the formation of black hole in failed CCSN. 
The detailed characteristics of neutrino emission depend not only on the properties of progenitor star (e.g., mass, compactness and so on~\cite{O'Connor_2013,Seadrow:2018ftp}), but also on the nuclear equation of state of neutron star which still remains largely uncertain~\cite{Pan:2017tpk,daSilvaSchneider:2020ddu,Raduta:2021coc}.
Except that, currently our comprehension on the spectral structure of supernova neutrinos is primarily obtained from studies on numerical simulations, due to lack of experimental data.

According to detailed investigations on supernova neutrino spectra~\cite{Keil:2002in,Tamborra:2012ac}, the instantaneous spectrum for each type of neutrinos will generally follow the quasi-thermal distribution (also called Garching formula), which can be presented as
\begin{equation}
\label{eq:vSpectrum}
    f_{\nu}(E_\nu) = \mathcal{A} \left(\dfrac{E_\nu}{\left< E_\nu \right>}\right)^\alpha \rm{exp}\left[ - (\alpha +1) \dfrac{E_\nu}{\left< E_\nu \right>} \right].
\end{equation}
Here, $E_\nu$ and $\left< E_\nu \right>$ are the energy and average energy of neutrino in the unit of MeV, respectively; $\mathcal{A} = \dfrac{(\alpha+1)^{\alpha+1}}{\left< E_\nu \right>\Gamma(\alpha+1)}$ is the normalization factor with $\Gamma$ being the gamma function;  and $\alpha$ characterises the amount of spectral pinching (with large value leading to suppression on high energy tail). 
$\alpha$ can be determined by the energy moment of the distribution, e.g., the relation
\begin{equation}
    \dfrac{\left< E_\nu^2 \right>}{\left< E_\nu \right>^2} =  \dfrac{2 + \alpha}{1 + \alpha}.
\end{equation}
Actually, eq.~\eqref{eq:vSpectrum} has been usually adopted as well to describe the time-integrated spectra in previous studies~\cite{Minakata:2008nc,Dasgupta:2011wg,Lu:2016ipr,Nikrant:2017nya,GalloRosso:2017mdz,GalloRosso:2020qqa}, and so do we.
Now, assuming no neutrino oscillation, the flux on the Earth can be expressed as
\begin{equation}
\label{eq:vFluence}
    \Phi(E_\nu) = \dfrac{1}{4 \pi d^2} \dfrac{\mathcal{E}_\nu}{\left< E_\nu \right>} f_{\nu}(E_\nu),
\end{equation}
where $d$ is the distance of source, and $\mathcal{E}_\nu$ denotes the total energy emitted through a specific species of neutrinos.
The spectral parameters for the source, adopted in this work, are given in table~\ref{tab:spectral_parameters}.
\begin{table}[htbp]
  \caption{Spectral parameters for the time-integrated spectra of supernova neutrino fluxes (see table 1 in ref.~\cite{Nikrant:2017nya}).}
  \label{tab:spectral_parameters}
  \centering
  \begin{tabular}{lccc} \hline
        $\nu$ & $\alpha_\nu$ & $\left< E_{\nu} \right>$ [MeV] & $\mathcal{E}_\nu$ [$10^{52}$ erg]\\ \hline
    $\nu_e$    & 2.67  & 14.1 & 7.70 \\
    $\bar{\nu}_e$ & 3.28 & 16.3 & 6.44  \\
    $\nu_x$ & 2.20 & 17.2 & 5.88 \\
    \hline
  \end{tabular}
\end{table}
It should be mentioned that the progenitor model, used to generate these parameters in the simulation, is expected to explode as one of the most common type II supernova (see ref.~\cite{Nikrant:2017nya} for more details).

\begin{figure}[htbp]
     \centering
     \begin{subfigure}[b]{0.45\textwidth}
         \centering
         \includegraphics[width=\textwidth]{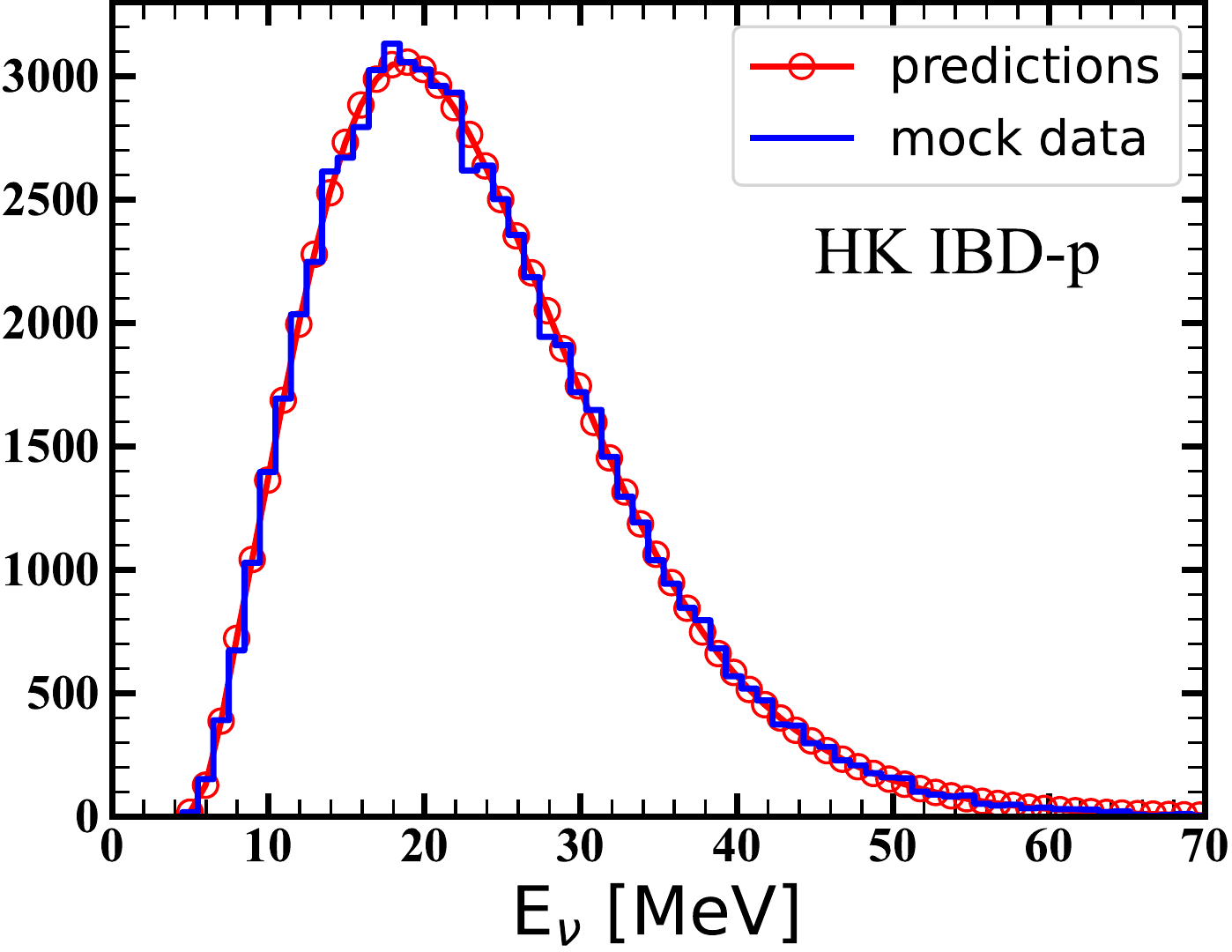}
         \caption{IBD-p in Hyper-K.}
         \label{subfig:HKibd}
     \end{subfigure}
     \hfill
     \begin{subfigure}[b]{0.45\textwidth}
         \centering
         \includegraphics[width=\textwidth]{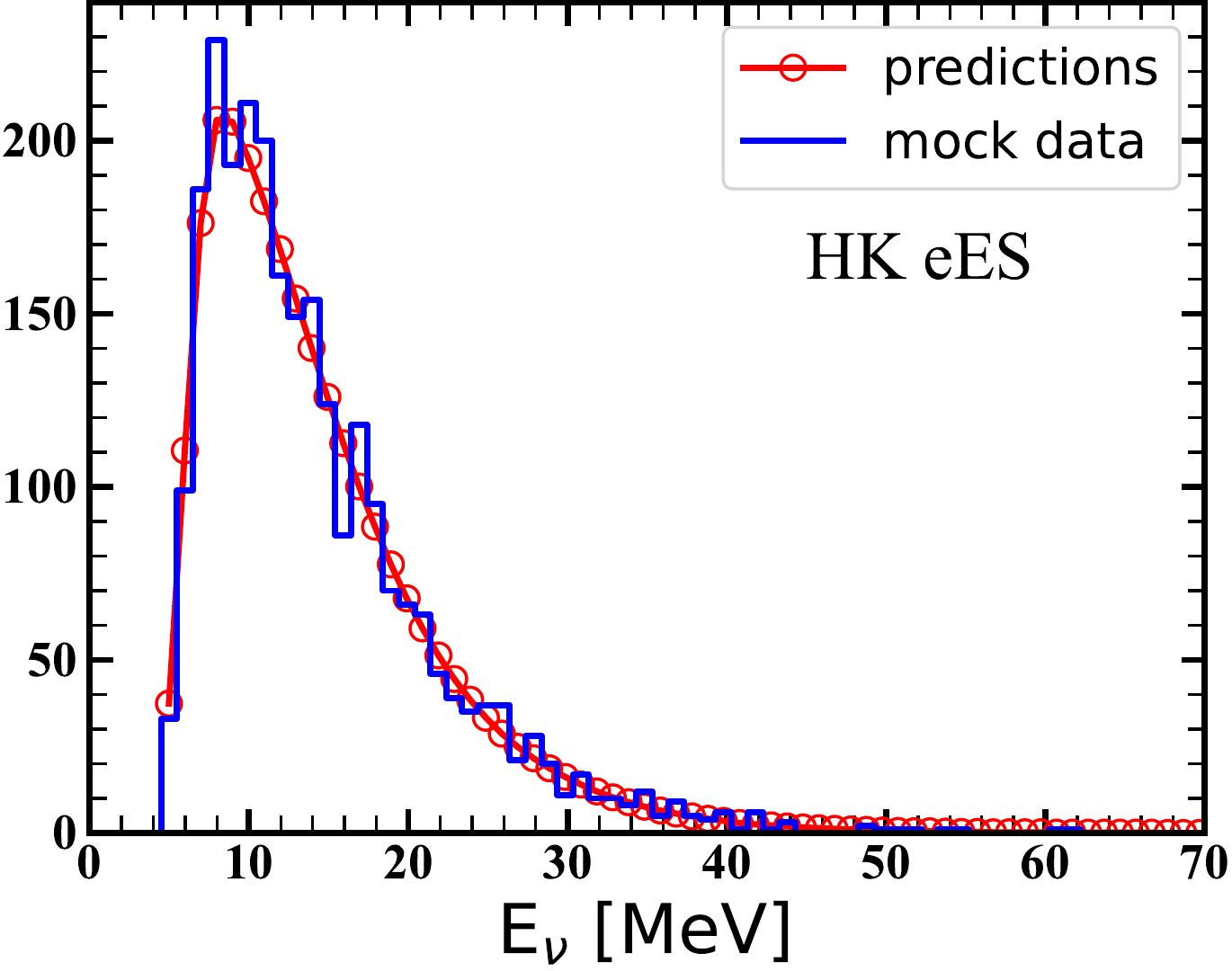}
         \caption{eES in Hyper-K.}
         \label{subfig:HKeES}
     \end{subfigure}
     \hfill
     \begin{subfigure}[b]{0.45\textwidth}
         \centering
         \includegraphics[width=\textwidth]{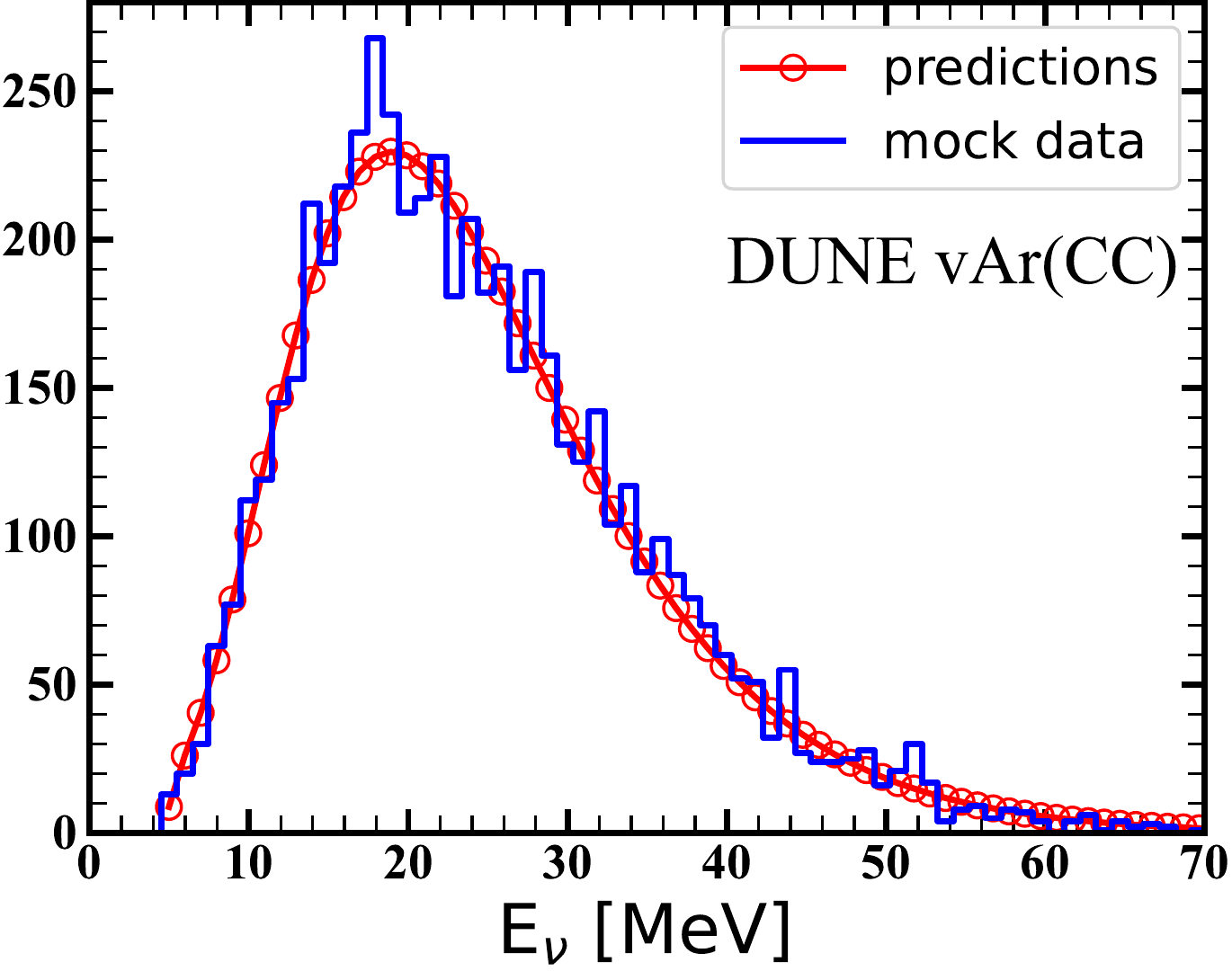}
         \caption{vAr(CC) in DUNE.}
         \label{subfig:DUNEvArCC}
     \end{subfigure}
     \hfill
     \begin{subfigure}[b]{0.45\textwidth}
         \centering
         \includegraphics[width=\textwidth]{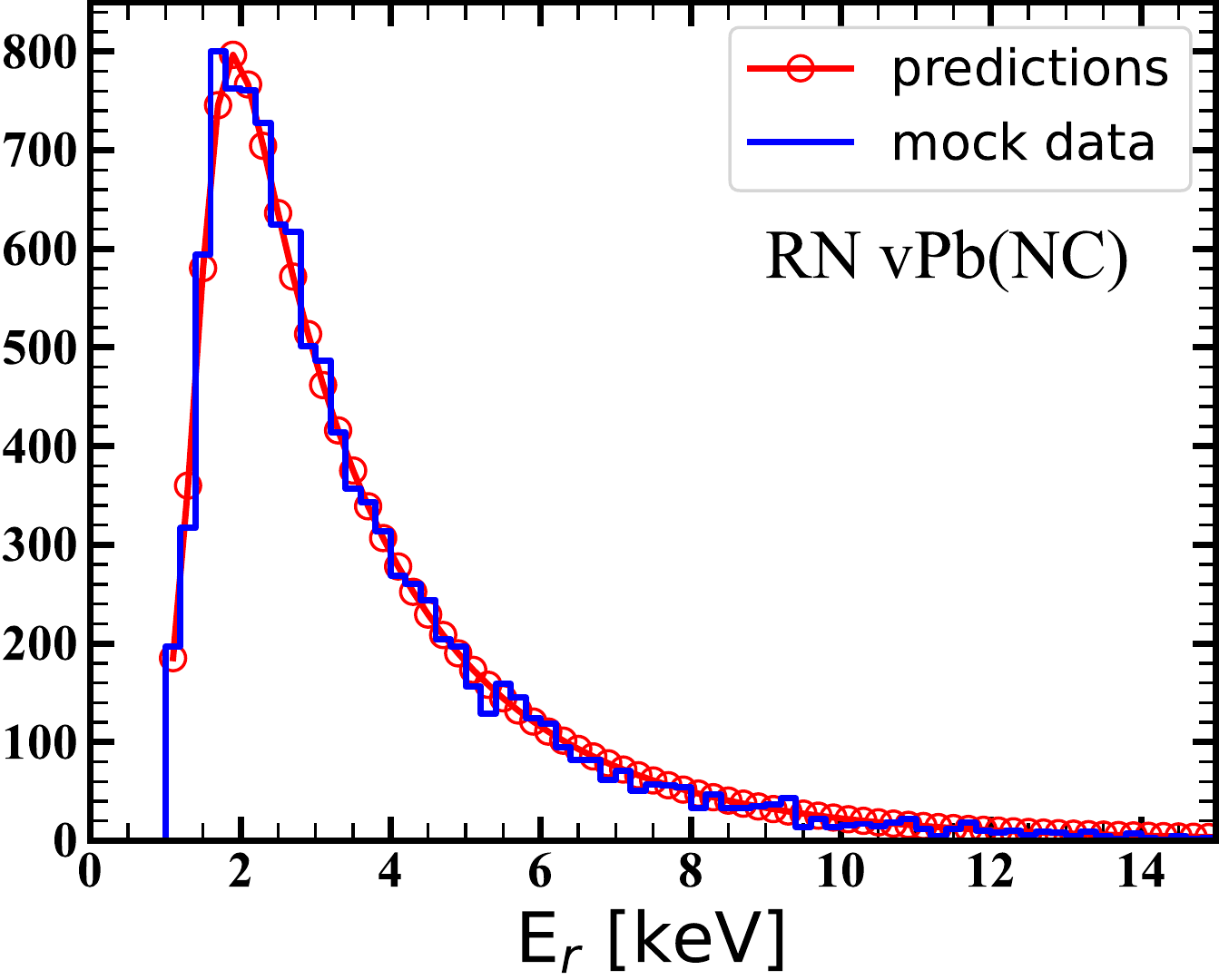}
         \caption{vPb(NC) in RES-NOVA.}
         \label{subfig:RNvPbNC}
     \end{subfigure}
        \caption{Predicted events and mock data for each reaction channel in Hyper-k, DUNE and RES-NOVA. $E_\nu$ and $E_r$ are the reconstructed neutrino energy and nuclear recoil energy, respectively. The source is assumed to be located at a typical distance, i.e., $d = 10 \kpc$, and no oscillation effect is under evaluation.}
        \label{fig:pred&mock}
\end{figure}

Now, the predicted event rate for each channel can be calculated.
For Hyper-K and DUNE, we set a uniform 100 energy grids to cover the energy range of $0.25 - 100.00 \MeV$,~\footnote{A uniform energy grid of 200 bins has been fixed as the highest energy resolution in \textit{SNOwGLoBES}, which covers the energy range of $0.25 - 100.00 \MeV$.} and drop the first several bins to approximately obtain a threshold of $5 \MeV$.
For RES-NOVA, we also set a uniform energy grid with the bin width of $0.2 \keV$, which starts from the threshold of $1 \keV$~\footnote{Such arrangement corresponds to a non-uniform energy grid on neutrino energy with the largest bin width being $0.94 \MeV$.}. 
We have also tested another non-uniform grid scheme, i.e., the adaptive energy-gridding technique~\footnote{In practice, we modify the adaptive energy-gridding technique, especially in the low-energy region, before it is implemented in our analysis. To be more specific, the similar strategy, used to handle the energies above the peak, has been applied to deal with the energies below the peak. Because we believe a proper treatment of the low energy part is quite necessary in order to achieve a better precision.} (see ref.~\cite{Nagakura:2020bbw}), and the results of analysis turn out to be almost the same as that of current grid scheme.
With the prediction data, the mock data can be generated now, e.g., given the predicted number of events $N_{pd}$, the corresponding number $N_{md}$ can be extracted from a Poisson distribution with $N_{pd}$ being the average value~\footnote{Such strategy has already been used in previous works, e.g., see ref.~\cite{GalloRosso:2017mdz}.}. 
The results are shown in figure~\ref{fig:pred&mock}.
The caveat is that such a treatment means that the mock data is extracted from one simulated measurement. 
So, it is inevitable that the information reflected by the data may deviates from that of the original source due to the Poisson processes.
Only high statistics can alleviate such deviations.
However, this is also the fact faced by realistic measurements.

Flavour transitions are also inevitable for supernova neutrinos.
These messengers are primarily produced in the dense core of a dying star, penetrate through the thick stellar mantle and ultimately arrive in detectors on the Earth.
Various conditions, encountered in this long journey, lead to complex transition patterns, e.g., adiabatic/non-adiabatic transitions, self-induced transitions and earth matter effects~\cite{Mirizzi:2015eza,Liao:2016uis,Scholberg:2017czd}.
Since this work is not meant to dig into the detail of flavour conversion, we focus on the adiabatic transition associated with smoothly-varying matter potentials in supernovae, for simplicity.
On the other hand, the three-flavour neutrino mixing framework has been well established experimentally due to tremendous experimental efforts over the past few decades.
So we can describe the flavour transitions in supernovae with proper formulas under specific assumptions.
However, there still exist two unknowns up to now in this scenario, i.e., the mass ordering and the complex phase associated with CP-violating observable.
For the latter one, previous works have shown that it will not cause sizeable modifications to the signals of supernova neutrinos~\cite{Akhmedov:2002zj,Balantekin:2007es}.
But the previous one is crucial to the flavour composition of supernova neutrinos in detectors.
And that necessitates the consideration of both NMO and IMO in this work.

Assuming the adiabatic Mikheyev-Smirnov-Wolfenstein (MSW) model, in the case of NMO, the observed fluxes ($\Phi_{\nu}$) are composed with the original fluxes ($\Phi_{\nu}^0$) in the following forms~\cite{Scholberg:2017czd}:
\begin{align}
    \Phi_{\nu_e} &= \Phi^0_{\nu_x} \qquad \qquad \qquad \qquad \qquad \quad  (\mathrm{NMO}), \label{eq:NH_ve} \\
    \Phi_{\bar{\nu}_e} &= \cos^2\theta_{12} \Phi^0_{\bar{\nu}_e} + \sin^2\theta_{12} \Phi^0_{\bar{\nu}_x}  \qquad (\mathrm{NMO}), \label{eq:NH_vebar}
\end{align}
where $\theta_{12}$ is the mixing angle with the value $\sin^2 \theta_{12} = 0.307 \pm 0.013$~\cite{ParticleDataGroup:2022pth}.
In the case of IMO, the formulas are rearranged as~\cite{Scholberg:2017czd}
\begin{align}
    \Phi_{\nu_e} &= \sin^2\theta_{12} \Phi^0_{\nu_e} + \cos^2\theta_{12} \Phi^0_{\nu_x}  \qquad (\mathrm{IMO}), \label{eq:IH_ve}\\
    \Phi_{\bar{\nu}_e} &= \Phi^0_{\bar{\nu}_x} \qquad \qquad \qquad \qquad \qquad \quad  (\mathrm{IMO}). \label{eq:IH_vebar}
\end{align}
And the total fluxes are conserved in both cases with such an equality: 
\begin{align}
    \Phi_{\nu_e} + \Phi_{\bar{\nu}_e} + 4 \Phi_{\nu_x} = \Phi^0_{\nu_e} + \Phi^0_{\bar{\nu}_e} + 4 \Phi^0_{\nu_x} \qquad (\mathrm{NMO \& IMO}).
\end{align}
Here $\Phi_{\nu_x}$ and $\Phi_{\bar{\nu}_x}$ represent the fluxes of neutrinos and anti-neutrinos with heavy flavours, sequentially, and are all equal to one quarter of the total heavy flavour flux.
In the data analyses, we do not distinguish between them.
From the above expressions, one can see that in the NMO case, the $\nu_e$ component is ultimately coming from the original $\nu_x$ component while the $\bar{\nu}_e$ flavour is only partially transformed.
In the IMO case, the transformations is almost reversed, i.e., the $\bar{\nu}_e$ flavour is fully transformed now while the $\nu_e$ component is partially transformed.
Note that, instead of simply reversion, the extents of partial transformations are different for the two cases. 

\begin{figure}[htbp]
     \centering
     \begin{subfigure}[b]{0.45\textwidth}
         \centering
         \includegraphics[width=\textwidth]{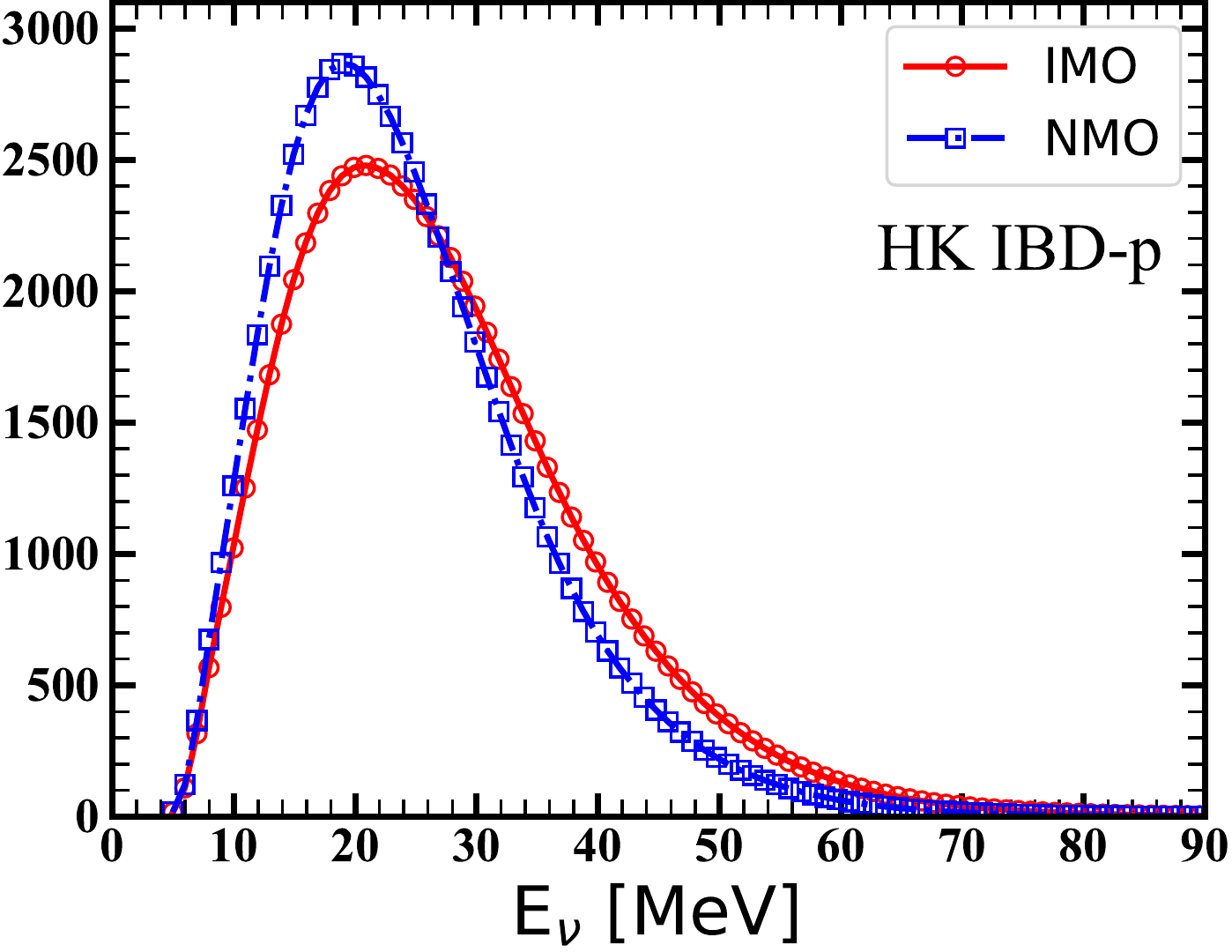}
         \caption{IBD-p in Hyper-K.}
         \label{subfig:HKibd&osc}
     \end{subfigure}
     \hfill
     \begin{subfigure}[b]{0.45\textwidth}
         \centering
         \includegraphics[width=\textwidth]{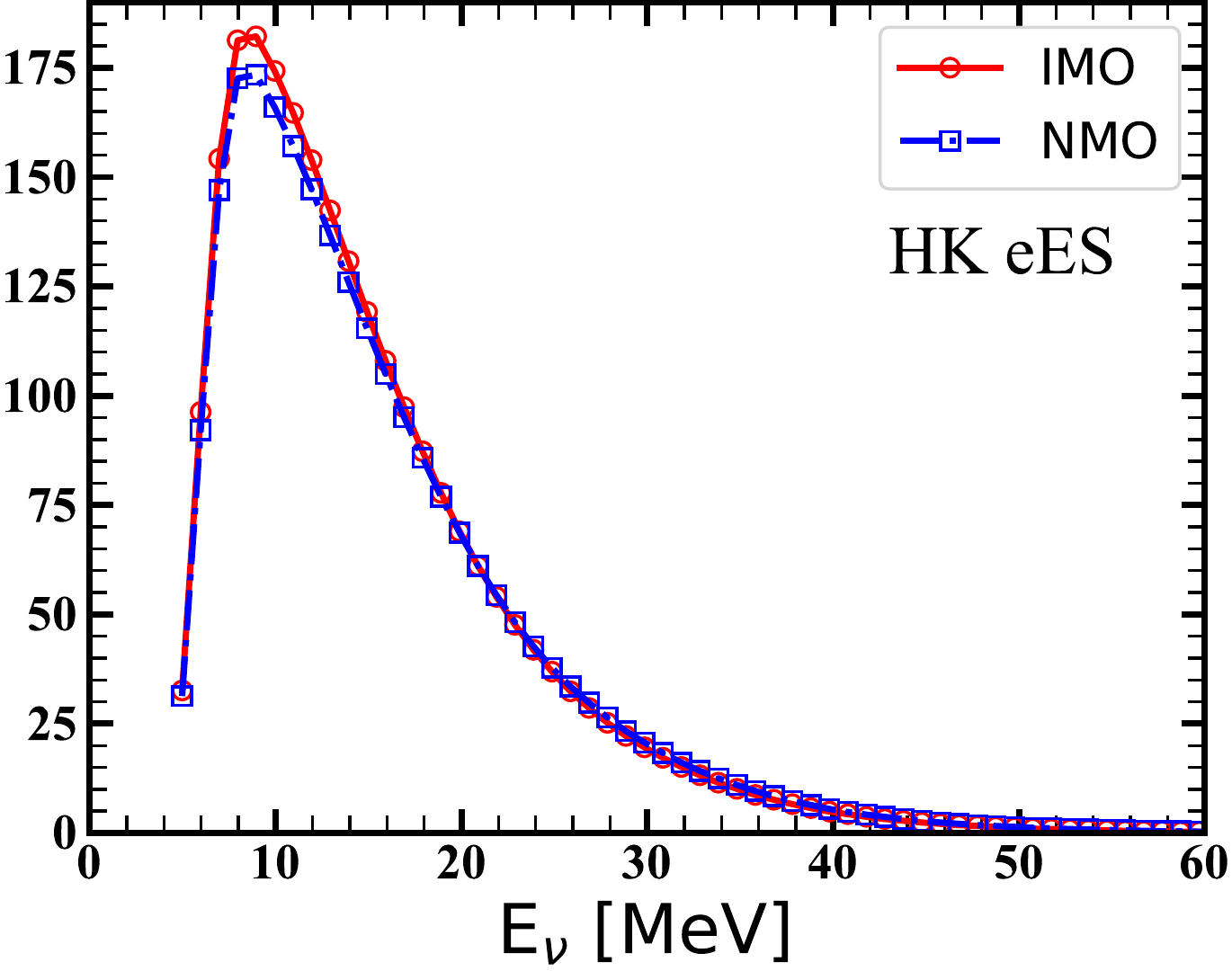}
         \caption{eES in Hyper-K.}
         \label{subfig:HKeES&osc}
     \end{subfigure}
     \hfill
     \begin{subfigure}[b]{0.45\textwidth}
         \centering
         \includegraphics[width=\textwidth]{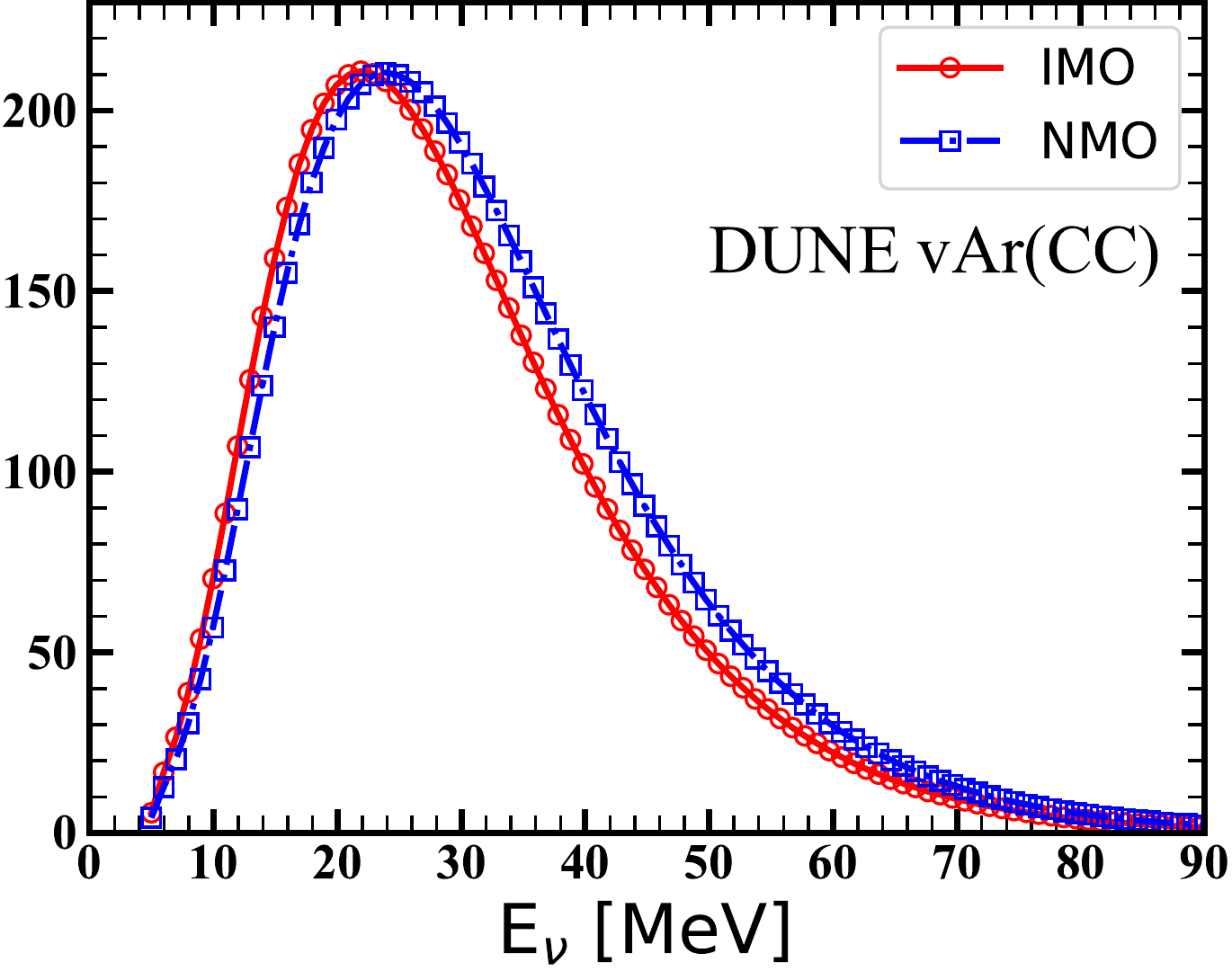}
         \caption{vAr(CC) in DUNE.}
         \label{subfig:DUNEvArCC&osc}
     \end{subfigure}
     \hfill
     \begin{subfigure}[b]{0.45\textwidth}
         \centering
         \includegraphics[width=\textwidth]{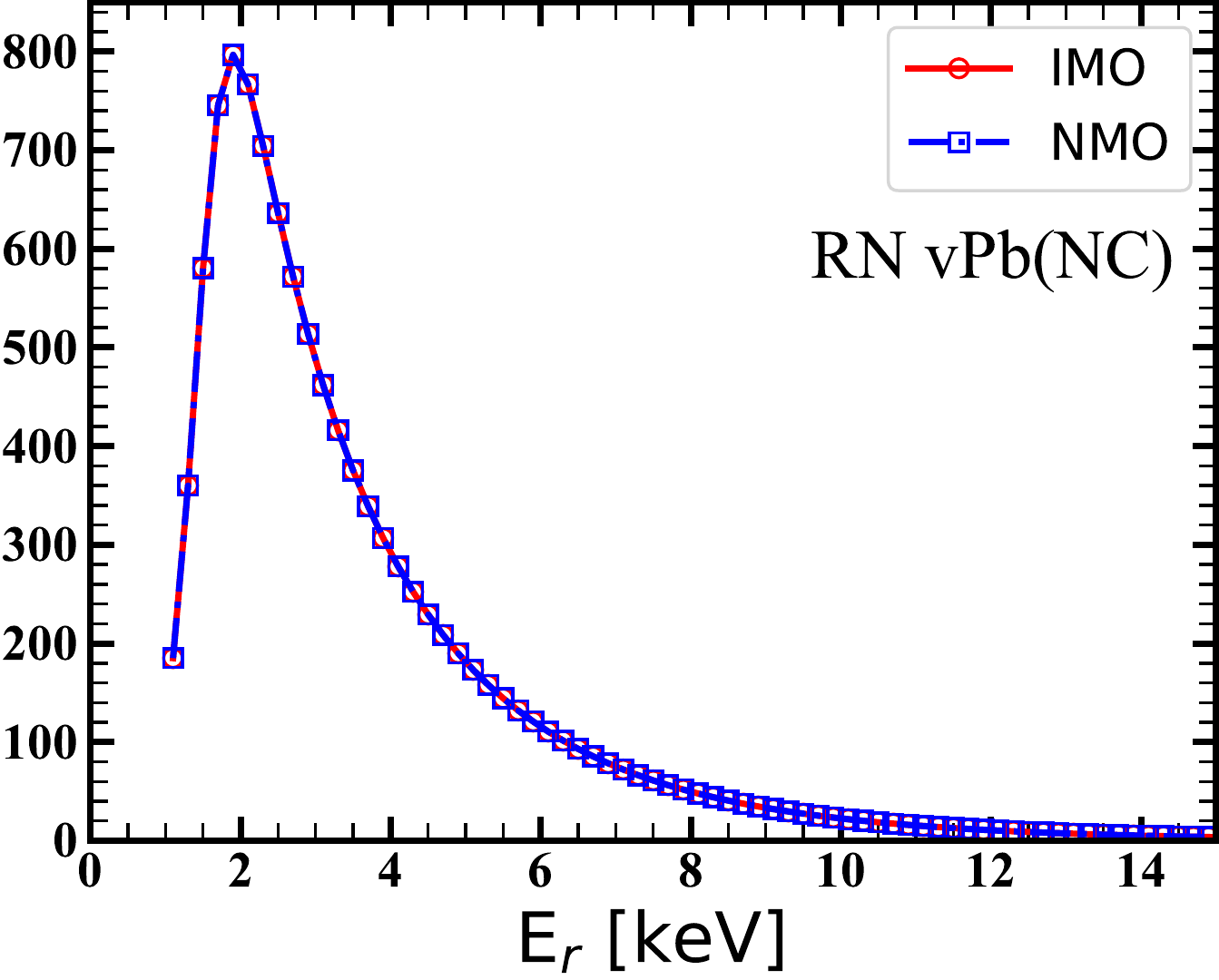}
         \caption{vPb(NC) in RES-NOVA.}
         \label{subfig:RNvPbNC&osc}
     \end{subfigure}
        \caption{Predicted events in each reaction channel under inverted mass ordering (IMO) or normal mass ordering (NMO). $E_\nu$ ($E_r$) denotes the reconstructed neutrino energy (nuclear recoil energy). The distance is assumed to be 10 kpc. The mock data for each case can be extracted with the same strategy in figure~\ref{fig:pred&mock} and we did not show them here.}
        \label{fig:pred&osc}
\end{figure}

The oscillation effects on the prediction of each reaction channel are shown in figure~\ref{fig:pred&osc}.
As one can see, it is clear that the predicted energy spectra for different mass ordering diverge from each other in the flavour-sensitive reaction channels, including IBD-p and eES in Hyper-K and vAr(CC) in DUNE, while they totally overlap with each other in the flavour-blind reaction channel, i.e., vPb(NC) in RES-NOVA.
It is also interesting to mention that the different gaps between IMO and NMO in IBD-p and vAr(CC) reflect the different extents of partial transformations.
For the mock data used in the final analysis, we conduct the same extractions, only including  all those ingredients this time.

\section{Bayesian inference and numerical results}
\label{sec:BayInf&NumRes}

Now data analysis can be performed with Bayesian inference to the mock data generated in the previous section.
We firstly describe the basic ideas of Bayesian inference briefly and the prior arrangements of our analysis. 
Then what's following are the demonstration of numerical results and some discussions as well.

\subsection{Basic ideas}

Bayesian statistics is fundamentally different from conventional frequentist statistics.
In Bayesian probability theory, probability is treated as a subjective concept which depends on our state of knowledge, instead of the objective limit of relative frequency of the outcome.
So, it is allowed to get updated on the basis of new information which can be collected via some approaches, e.g., conducting experiments.
With a full understanding of the issue under investigation, in principle the Bayesian probability will arrive at a stable value.
The basic logical rule which allows us to do such updating is the Bayes' theorem, which can be presented as
\begin{equation}
\label{eq:BayesTheorem}
    P(\theta|D) \propto P(D|\theta) P(\theta).
\end{equation}
In the case of parameter estimation, $\theta$ and $D$ represent the model parameter to be estimated collectively and the dataset relevant to the model, respectively.
The quantity to be evaluated is the posterior probability, $P(\theta|D)$, which stands for the probability of $\theta$ given the new dataset $D$.
$P(\theta)$ is the prior probability which quantifies our beliefs on $\theta$ before inclusion of new conditions. 
The likelihood function $P(D|\theta)$ is a mathematical function of $\theta$ for a fixed dataset $D$ (also denoted by $\mathcal{L}(\theta;D)$).
It quantifies the probability of the observation of $D$ when given the specific parameter $\theta$.
In this framework, the main task of inference will get descended into how to calculate the distribution of posterior probability, once the expressions of prior and likelihood are settled.
Note that a proper realization of prior probability will be quite helpful in the analysis of less informative dataset, but, somehow, trivial in the case with dataset informative enough.

In this work, since the Garching formula is adopted to describe the time-integrated spectra of supernova neutrinos, we get 9 model parameters, i.e.,
\begin{equation}
    \Vec{\theta} = (\alpha_{\nu_e}, \alpha_{\bar{\nu}_e}, \alpha_{\nu_x}, \left< E_{\nu_e} \right>, \left< E_{\bar{\nu}_e} \right>, \left< E_{\nu_x} \right>, \mathcal{E}_{\nu_e}, \mathcal{E}_{\bar{\nu}_e}, \mathcal{E}_{\nu_x}).
\end{equation}
The realisation of $P(\Vec{\theta})$ could be nontrivial.
Generally speaking, the posterior distribution of previous inference can act as the prior distribution of new inference with new information.
However, this is not the case in this study, due to the highly limited information provided by the measurement of SN 1987A.
Up to now, our knowledge on this issue is primarily obtained from various simulations.
In detail, the values of $\alpha$ are usually varying with time in the range of $2 \lesssim \alpha \lesssim 4$~\cite{Hudepohl:2009tyy,Keil:2002in,Tamborra:2012ac}.
For $\left< E_{\nu} \right>$, the magnitude of $\sim 10 \MeV$ exists in almost all simulations and also gets confirmed by the observation of SN 1987A. Furthermore, a neutrino energy hierarchy is emerged as $\left< E_{\nu_e} \right> < \left< E_{\bar{\nu}_e} \right> \lesssim \left< E_{\nu_x} \right>$ in simulations~\cite{Hudepohl:2009tyy,Janka:2012wk}.
For $\mathcal{E}_{\nu}$, both simulations and SN 1987A indicate that the total released energy via neutrinos should lie in the vicinity of $3 \times 10^{53} \erg$.
And the ansatz of energy equipartition among different flavours of neutrinos has also been found to be roughly valid in simulations.
Based on the above statements, we quantify the prior knowledge with 9 independent Gaussian functions associated with the 9 spectral parameters, i.e.,
\begin{equation}
\label{eq:prior_gauss}
    \log P(\theta > 0) = -\dfrac{(\theta - \mu)^2}{2 \sigma^2} + constant,
\end{equation}
where we exclude the non-physical negative quadrants.
The relevant Gaussian parameters are given in table~\ref{tab:prior_gauss}.
\begin{table}[htbp]
  \caption{The parameters of Gauss distributions in priors. $\mu$ and $\sigma$ represent the center values and standard deviations, respectively. $\left< E_{\nu} \right>$s are in the unit of MeV.}
  \label{tab:prior_gauss}
  \centering
  \begin{tabular}{lccccc} \hline
           & $\alpha_\nu$ & $\left< E_{\nu_e} \right>$ & $\left< E_{\bar{\nu}_e} \right>$ & $\left< E_{\nu_x} \right>$ & $\mathcal{E}_\nu$ [$10^{52}\erg$]\\ \hline
    $\mu$    & 3  & 12 & 14 & 16 & 5\\
    $\sigma$ & 1 & 4 & 4 & 4 & 5/3  \\ \hline
  \end{tabular}
\end{table}
It must be emphasized here that, with such arrangements, we do not intend to mean that the spectral parameters of neutrinos from the next galactic CCSN would follow these distributions.
It rather expresses such a belief that we are quite confident that $\theta$ will lie within $\mu \pm \sigma$, very sure that $\theta$ will lie within $\mu \pm 2\sigma$ and almost certain that $\theta$ will lie within $\mu \pm 3\sigma$.
Values far beyond these regions are still possible but just not likely to happen since that would break the current theoretical framework.
Such priors cover the parameter spaces used in the previous analysis~\cite{GalloRosso:2017mdz} with the regions of $3\sigma$, and meanwhile accommodate strong deviations from the expected values.
However, it should be noted again that the posterior will be eventually dominated by the data, instead of the choice of priors, when the dataset is informative enough.
As a confirmation, we also conduct the analysis with flat priors, and the comparison is shown in appendix~\ref{apd:priors}.

The dataset consists of series of energy bins and related number of events, and we conduct the analysis with such a binned likelihood:
\begin{equation}
    \label{eq:likelihood}
    \mathcal{L}_{\zeta}(\Vec{\theta};D) = \prod_{i=1}^{\mathrm{N}_{\mathrm{bin}}} \dfrac{\lambda_i^{n_i}}{n_i !} \ee^{-\lambda_i},
\end{equation}
for the reaction channel $\zeta$, where $\mathrm{N}_{\mathrm{bin}}$ is the number of energy bins, $\lambda_i$ and $n_i$ represent the number of events related to the $i$th bin in predictions and mock data, respectively.
$\lambda_i$ is a function of $\Vec{\theta}$, while $n_i$ belongs to $D$.
Such a Poisson distribution is also adopted in previous studies~\cite{GalloRosso:2017mdz,GalloRosso:2020qqa}.
Now, the eventual likelihood is simply expressed as
\begin{equation}
    \mathcal{L}(\Vec{\theta};D) = \prod_{\zeta \in \ all \ exp.} \mathcal{L}_{\zeta}(\Vec{\theta};D),
\end{equation}
after combining all the reaction channels.
Other potentially useful reaction channels can also be considered via this formula in the future.
Furthermore, eq.~\eqref{eq:likelihood} can be replaced with another more well-constructed likelihood, which considers other uncertainties in realistic measurements thoroughly, in future studies.

The calculation of posterior distribution used to be the most complicated part of Bayesian inference.
However, powerful methods and tools are currently available to alleviate it. 
In this work, we implement the ensemble sampler tool, \textit{emcee}~\footnote{\url{https://emcee.readthedocs.io/en/stable/index.html}.}~\cite{Foreman-Mackey:2012any}, to sample the 9-dimension posterior distribution.
The \textit{emcee} package is a Python implementation of the Metropolis–Hastings (M–H) algorithm~\footnote{The M-H algorithm is the most commonly used Markov chain Monte Carlo algorithm (see ref.~\cite{DAgostini:2003bpu,Foreman-Mackey:2012any} for more details).}, which has already been adopted in many published projects in the astrophysics literature.
The caveat, which derives from the M-H algorithm, is that the samples initially generated in the chain can be heavily influenced by the choice of starting point in parameter space, due to the inevitably existing correlation among neighboring samples.
So, in practice the initial part will be excluded.
In this study, we just drop the initial 200 samples in each chain, to obtain sets of stable samples.

\subsection{Demonstration}

Finally, we perform the analysis and show the numerical results in this section.
As a start, we test the capability of our method with the case considering no oscillation effects.
So as to obtain an appropriate determination of the posterior distribution, we draw a dataset including $10^6$ samples, and calculate the distribution of each parameter by conducting marginalization over other parameters which follows the law of total probability.
Then, the cases of different oscillation models are evaluated through the same processes.
Among them, $d = 10 \kpc$ is adopted as the default distance of source.
In practice, it is possible for this distance to be much smaller, e.g., nearby core collapse supernova candidates reported in \cite{Mukhopadhyay:2020ubs}, including the famous Betelgeuse.
Generally speaking, smaller distance means higher statistics and then better precision, when neutrino flux is not too intense to cause signal pile-up in the detector.
However, it would be another topic, for future works, on how to properly deal with the effects of signal pile-up if the source is too close. 
As a test, we only estimate the distance effect in the case of $d = 5 \kpc$ here.

\subsubsection{No oscillation}

\begin{figure}[htbp]
    \includegraphics[width=\textwidth]{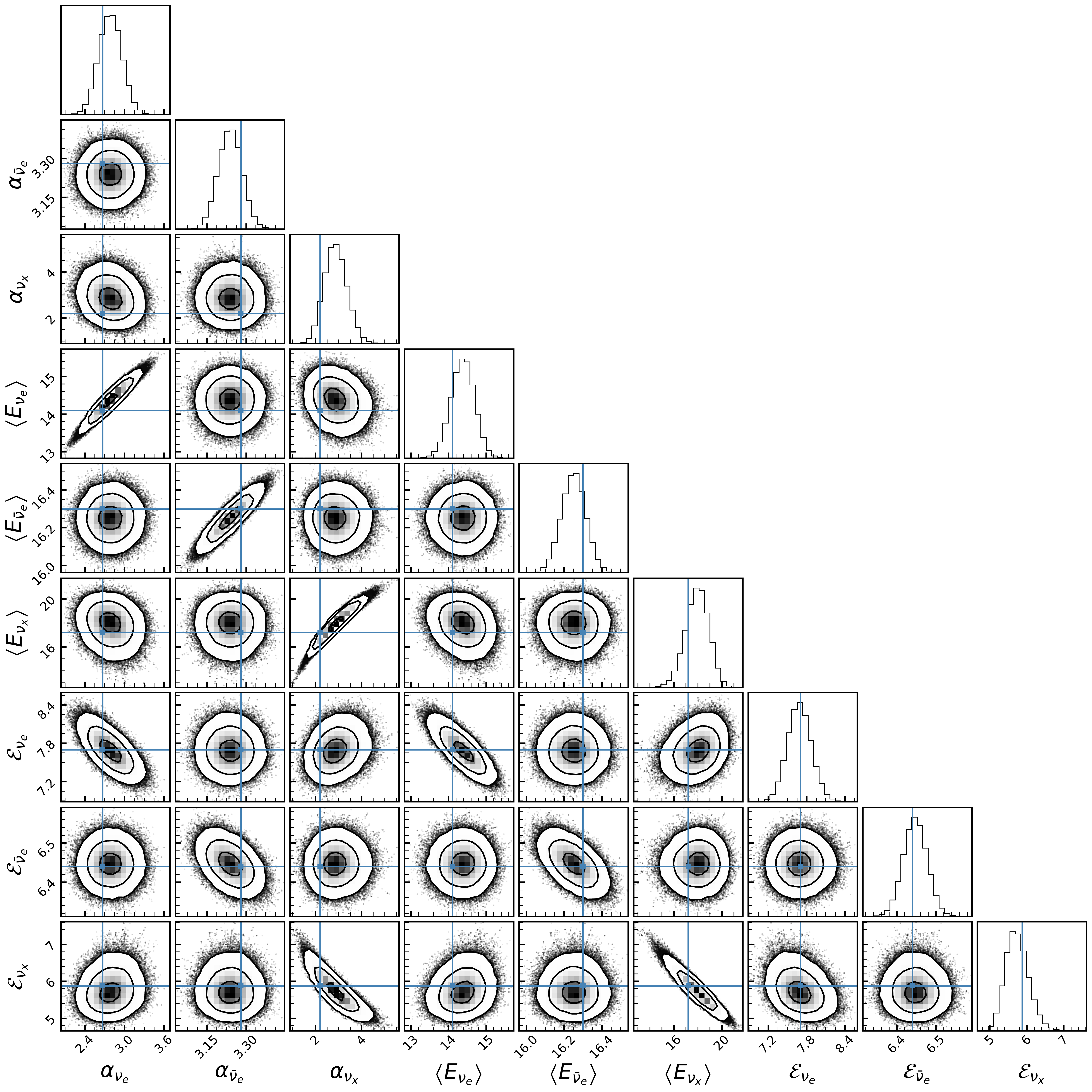}
    \caption{Posterior distributions for the no oscillation case. The Gaussian prior distributions are functioning here. Plots on the diagonal show posterior distributions for the corresponding parameter after marginalization over other parameters, and the off-diagonal ones show correlations between them. Contours in the off-diagonal plots demonstrate the area of $1 \sigma, 2 \sigma$ and $3 \sigma$ credible level, respectively. The blue lines mark the parameter values to generate the mock data used in this analysis.}
    \label{fig:bayfit_NO_gauss}
\end{figure}


Figure~\ref{fig:bayfit_NO_gauss} shows the posterior distributions when no neutrino oscillation is considered\footnote{\textit{corner} is used to plot such diagrams~\cite{corner}.}.
The 1-dimension (1-D) distributions for all spectral parameters are plotted on the diagonal. 
We also present the representative values of these 1-D distributions, i.e., the maximum \textit{a posteriori} (MAP) estimate and the $2 \sigma$ credible intervals in the highest posterior density scheme, in table~\ref{tab:bayfit_results}.
As one can see, the three parameters for $\bar{\nu}_e$ flux are constrained quite well in this analysis.
In detail, the $2 \sigma$ symmetrized fractional uncertainties~\footnote{Indeed, asymmetries appear among these 1-D distributions in figure~\ref{fig:bayfit_NO_gauss} and also in table~\ref{tab:bayfit_results}, and will also show up in that of other cases. For simplicity, the symmetrized fractional uncertainties are calculated by averaging the positive and negative uncertainties over the most probable values here and after.} reach $\pm 2.8 \%$, $\pm 0.8 \%$ and $\pm 0.9 \%$ for $\alpha$, $\left< E \right>$ and $\mathcal{E}$, sequentially.
Such high precision is primarily attributed to the ultra-high statistics provided by the IBD-p channel in Hyper-K, as can be seen in figure~\ref{subfig:HKibd}.
Meanwhile, the sensitivity to $\nu_e$ mainly derives from the vAr(CC) reaction in DUNE and the eES channel in Hyper-K. 
Modest uncertainties are also achieved as $\pm 12.4 \%$, $\pm 4.3 \%$ and $\pm 4.9 \%$.
However, the precision for $\nu_x$ flux is relatively poor and the fractional uncertainties are only obtained as $\pm 33.4 \%$, $\pm 10.7 \%$ and $\pm 10.9 \%$.
The vPb(NC) reaction in RES-NOVA renders the primary sensitivity to $\nu_x$ and also achieves a number of total events even larger than the sum of that from the eES channels in Hyper-K and the vAr(CC) channel in DUNE.
However, the fact is that $\sim 1/3$ of the signals in RES-NOVA come from the $\nu_e$ and $\bar{\nu}_e$ fluxes.
That is, the information of $\nu_x$ from RES-NOVA is actually contaminated.
Nevertheless, higher statistics will further improve the accuracy, e.g., enlarging the fiducial mass of RES-NOVA by 10 times will improve the accuracy by $\sim 50\%$ in our test. 
On the other hand, due to the strong suppression of Pb nuclei on the nuclear recoil energy, a threshold of $1 \keV$ in nuclear recoil energy only makes RES-NOVA sensitive to neutrinos with energy above $\sim 10 \MeV$.
Such threshold, although literally quite low among detectors in the same category, is nevertheless not low enough for precision measurement of the spectrum of $\nu_x$ flux in supernova neutrinos, since the information below and even in the peak is lost.
Such loss naturally jeopardizes precision extraction of information related to spectral shape.~\footnote{In the test analyses, we assume a $\sim 6$ MeV threshold of neutrino energy (i.e., $0.4 \keV$ threshold of nuclear recoil energy) for RES-NOVA, and the accuracy for $\nu_x$ is improved by a factor of $1/4$. The neutral current scatterings on $^{16}\mathrm{O}$ in Hyper-K can also provide information on the low energy region (e.g., $\sim 400$ events in the energy range of $5 \sim 10 \MeV$). The inclusion of this reaction also lead to a moderate improvement ($\sim 25\%$) on the accuracy of $\alpha_{\nu_x}$.}

\begin{table}[htbp]
  \caption{The representative values of 1-D posterior distributions. NO indicates the case without neutrino oscillation, while NMO (IMO) represents the case of normal (inverted) mass ordering. Gaussian priors are adopted in all cases. The rows denoted with MAP give the most probable values of the posteriors, while $(2 \sigma^- , 2 \sigma^+)$ show the relative credible intervals at the $2 \sigma$ level of probability. $\%$ rows give the corresponding symmetrized fractional uncertainties. Meanwhile, $\%$(csu) rows show the estimated symmetrized fractional uncertainties after including a $\pm 5 \%$ uncertainty on the cross section of vAr(CC) reaction in DUNE.}
  \label{tab:bayfit_results}
  \centering
  \begin{threeparttable}[b]
     \begin{tabular}{ccccccccccc}
      \toprule
      \multirow{2}*{Osc} & \multirow{2}*{estimate} & \multicolumn{3}{c}{$\alpha$} & \multicolumn{3}{c}{$\left< E \right>$ [MeV]} & \multicolumn{3}{c}{$\mathcal{E}$ [$10^{52} \erg$]} \\
      \cmidrule(lr){3-5}\cmidrule(lr){6-8}\cmidrule(lr){9-11}
      & & $\nu_e$ & $\bar{\nu}_e$ & $\nu_x$ & $\nu_e$ & $\bar{\nu}_e$ & $\nu_x$ & $\nu_e$ & $\bar{\nu}_e$ & $\nu_x$  \\ 
      \midrule
      \multirow{4}{*}{NO} & MAP
      & 2.83 & 3.25 & 2.93 & 14.37 & 16.26 & 17.88 & 7.71 & 6.45 & 5.63 \\
      & $2 \sigma^-$ & -0.38 & -0.10 & -1.00 & -0.63 & -0.14 & -1.81 & -0.41 & -0.07 & -0.46 \\
      & $2 \sigma^+$ & +0.32 & +0.08 & +0.96 & +0.61 & +0.12 & +2.02 & +0.34 & +0.05 & +0.77 \\
      & $\%$ & 12.4 & 2.8 & 33.4 & 4.3 & 0.8 & 10.7 & 4.9 & 0.9 & 10.9 \\
      & $\%$(csu) & 13.3 & 2.8 & 38.6 & 4.7 & 0.8 & 12.9 & 8.9 & 1.0 & 14.7 \\
      \midrule
      \multirow{4}{*}{NMO} & MAP
      & 3.48 & 3.12 & 2.37 & 13.84 & 16.09 & 17.45 & 7.95 & 6.46 & 5.86 \\
      & $2 \sigma^-$ & -1.33 & -0.16 & -0.25 & -1.94 & -0.25 & -0.60 & -1.72 & -0.13 & +0.24 \\
      & $2 \sigma^+$ & +1.81 & +0.19 & +0.25 & +2.27 & +0.28 & +0.73 & +2.16 & +0.14 & +0.25 \\
      & $\%$ & 45.1 & 5.6 & 10.5 & 15.2 & 1.6 & 3.8 & 24.4 & 2.1 & 4.2 \\
      & $\%$(csu) & 47.3 & 6.2 & 11.8 & 20.4 & 1.8 & 4.3 & 34.6 & 3.9 & 8.4 \\
      \midrule
      \multirow{4}{*}{IMO} & MAP
      & 3.41 & 3.85 & 2.18 & 15.04 &  16.03 & 17.17 & 7.77 & 6.58 & 5.89 \\
      & $2 \sigma^-$ & -0.96 & -1.74 & -0.07 & -1.39 & -2.95 & -0.18 & -0.86 & -1.46 & -0.06 \\
      & $2 \sigma^+$ & +1.23 & +1.57 & +0.07 & +1.43 & +2.03 & +0.15 & +0.84 & +1.88 & +0.05 \\
      & $\%$ & 32.1 & 43.0 & 3.2 & 9.4 & 15.5 & 1.0 & 10.9 & 25.4 & 0.9 \\
      & $\%$(csu) & 34.4 & 46.1 & 3.7 & 12.0 & 19.3 & 1.0 & 20.5 & 35.9 & 0.9 \\
      \bottomrule
    \end{tabular}
  \end{threeparttable}
\end{table}

On the other hand, the off-diagonal plots suggest the correlations between parameters.
Generally speaking, it is quite noticeable that significant correlations appear among parameters in the same type of neutrinos universally, and also only exist among them.
Furthermore, these correlations even show certain features for a specific type of neutrinos, i.e., strong positive correlation between $\alpha$ and $\left< E \right>$ of whom both determine the shape of spectrum, and noteworthy negative correlations between $\mathcal{E}$ and one of the above spectral shape parameters, respectively.
Such correlation patterns are primarily embedded in the parameterization of neutrino spectrum (see eq.~\eqref{eq:vSpectrum} and eq.~\eqref{eq:vFluence}).
It is also potentially interesting to mention that such correlations are the weakest for the $\bar{\nu}_e$ flavour while that of the others are comparable~\footnote{We swap the parameters of $\nu_e$ and $\bar{\nu}_e$ components and such hierarchy still appears, only with the difference between these two species getting decreased by $\sim 50\%$. So it is primarily originated from the detection configurations.}.

The distance effect is tested here.
For a closer source with $d = 5 \kpc$, the higher statistics in data lead to better accuracies on the reconstructed spectral parameters, while almost no effect on the correlations among these parameters.
In detail, the symmetrized factional uncertainties are updated by $\pm 6.4 \%$, $\pm 2.3 \%$ and $\pm 2.6 \%$ for $\nu_e$ flavour, $\pm 1.5 \%$, $\pm 0.4 \%$ and $\pm 0.5 \%$ for $\bar{\nu}_e$ part and $\pm 20.1 \%$, $\pm 6.3 \%$ and $\pm 5.8 \%$ for $\nu_x$ component. However, as a result for comparison, these percentages are calculated with new $2 \sigma$ credible intervals (i.e., for $d = 5 \kpc$) and the most probable values in the previous case (i.e., for $d = 10 \kpc$).
Such treatment is also applied in similar comparisons hereafter.
In short, the accuracies are universally enhanced by $40\% \sim 50\%$ among all parameters in this test.

\subsubsection{Flavour conversions}


\begin{figure}[htbp]
	\includegraphics[width=\textwidth]{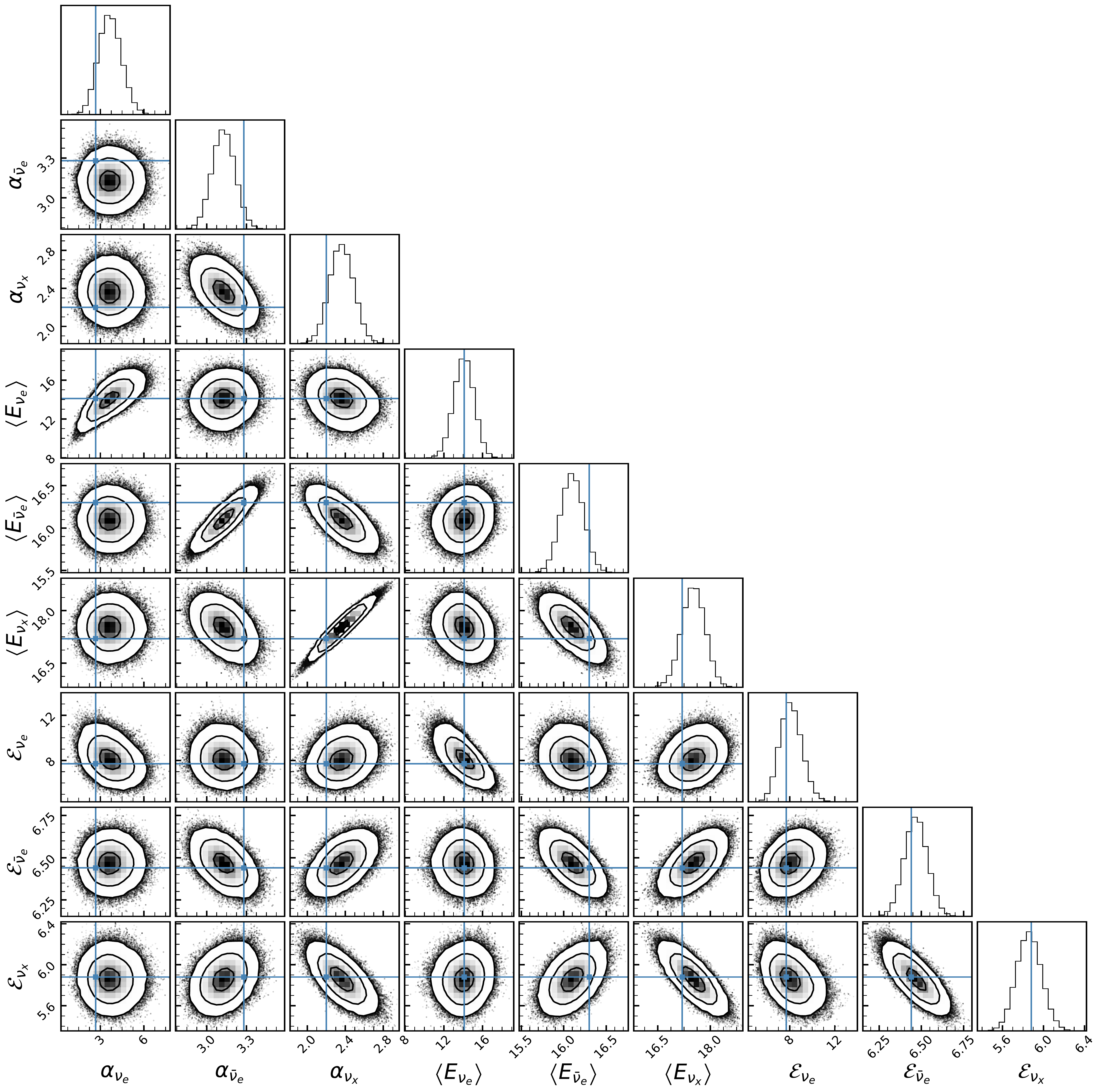}
    \caption{The same as figure~\ref{fig:bayfit_NO_gauss}, but the oscillation effects with normal mass ordering are under evaluation.}
    \label{fig:bayfit_NH_gauss}
\end{figure}

Figure~\ref{fig:bayfit_NH_gauss} displays the posterior distributions when the oscillation effects are considered under the assumption of NMO.
The representative values, corresponding to the distributions on the diagonal, are also given in table~\ref{tab:bayfit_results}.
Still, the best results are obtained for the $\bar{\nu}_e$ flavour for the same reason as the case without oscillation effect.
Numerically speaking, the symmetrized fractional uncertainties are $\pm 5.6 \%$, $\pm 1.6 \%$ and $\pm 2.1 \%$ for $\alpha$, $\left< E \right>$ and $\mathcal{E}$, sequentially, within a credible level of $2 \sigma$.
They become worse slightly, due to the partial conversion in eq.~\eqref{eq:NH_vebar}.
In this flavour conversion mode, the $\nu_e$ events and $\sim 30 \%$ of $\bar{\nu}_e$ events in detectors are now responsible for the $\nu_x$ component.
Thus, the results for $\nu_x$ component are much better after combining information from all the four channels.
The uncertainties are read as $\pm 10.5 \%$, $\pm 3.8 \%$ and $\pm 4.2 \%$, even slightly better than the $\nu_e$ results in the case of no oscillation.
In contrast, the precision for $\nu_e$ are now rather poor, only achieving uncertainties of $\pm 45.1 \%$, $\pm 15.2 \%$ and $\pm 24.4 \%$.
Because all the information for $\nu_e$ flavour are extracted from the data of vPb(NC) reactions in RES-NOVA, and only $\sim 1/6$ of these data are responsible.
Note that the deviation between posterior and prior distributions for $\alpha_{\nu_e}$ is kind of trivial, which means the result get too much information from the prior, instead of the data.
It indicates that the constraint on $\alpha_{\nu_e}$ is actually quite limited in this case.

\begin{figure}[htbp]
	\includegraphics[width=\textwidth]{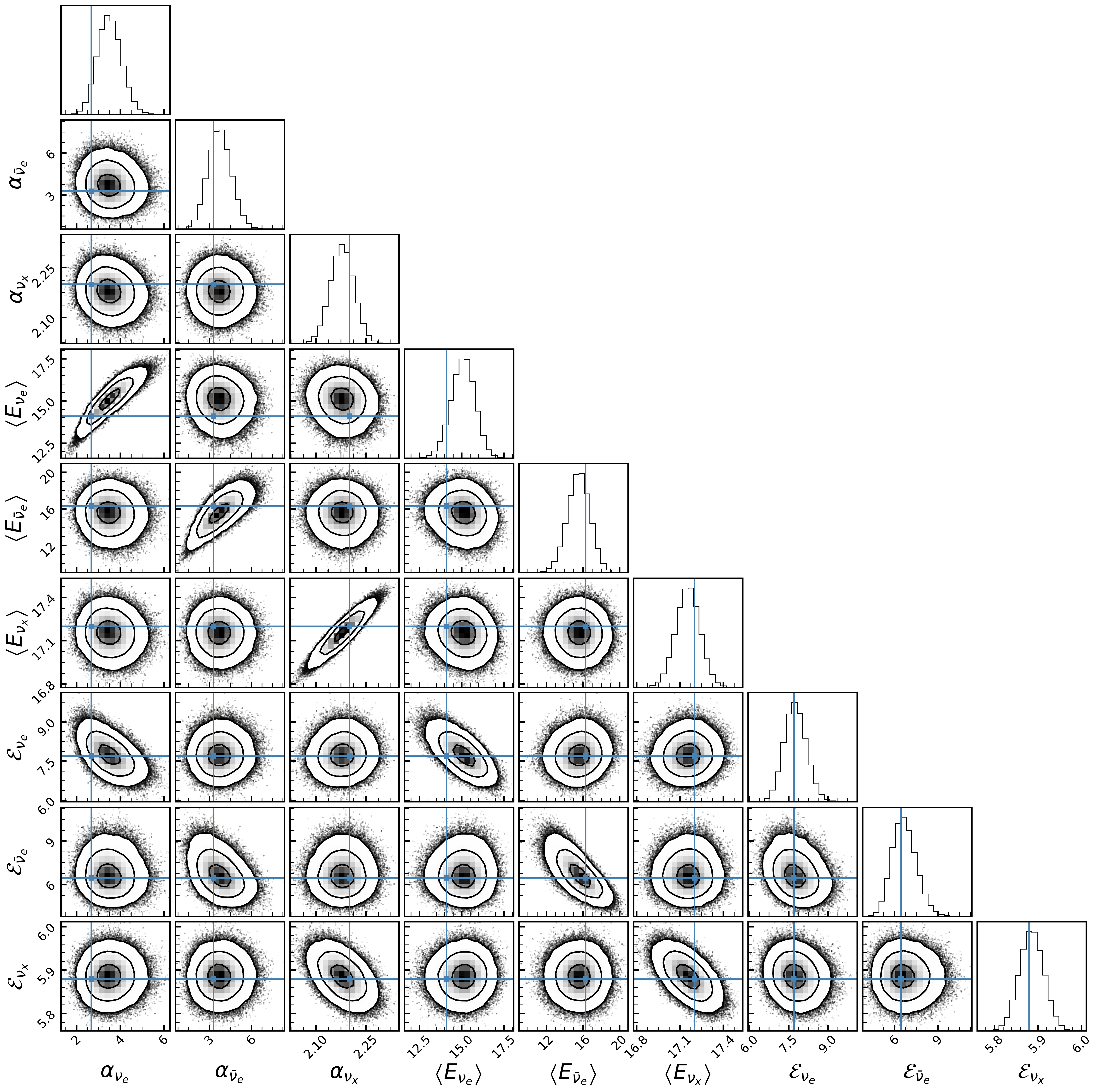}
    \caption{The same as figure~\ref{fig:bayfit_NO_gauss}, but the oscillation effects with inverted mass ordering are under evaluation.}
    \label{fig:bayfit_IH_gauss}
\end{figure}

The numerical results for the IMO conversion are illustrated in figure~\ref{fig:bayfit_IH_gauss} and table~\ref{tab:bayfit_results}.
In this conversion mode, neutrino signals in all reaction channels are mainly coming from the original $\nu_x$ component (see eq.~\eqref{eq:IH_ve} and eq.~\eqref{eq:IH_vebar}), which naturally lead to a promising precision in this part.
That is, the symmetrized fractional uncertainties are obtained as $\pm 3.2 \%$, $\pm 1.0 \%$ and $\pm 0.9 \%$ for the three parameters correspondingly.
It should be mentioned that $\nu_x$ components are responsible for $\sim 2/3$ of the total neutrinos.
Hence, it is quite significant to achieve such high precision on the measurement of this part.
However, the price is large uncertainties on the measurements of other components.
The representative values of posterior distributions are $\pm 32.1 \%$, $\pm 9.4 \%$ and $\pm 10.9 \%$ for the $\nu_e$ flavour, and $\pm 43.0 \%$, $\pm 15.5 \%$ and $\pm 25.4 \%$ for the $\bar{\nu}_e$ part.
The situation of the $\bar{\nu}_e$ part are quite similar to that of the $\nu_e$ flavour in the NMO conversion.
Similarly, the caveat is that the prior distribution provides too much information in the evaluation of $\alpha_{\bar{\nu}_e}$, also just like the case of $\alpha_{\nu_e}$ in the NMO conversion.

\begin{figure}[htbp]
     \centering
     \begin{subfigure}[b]{0.48\textwidth}
         \centering
         \includegraphics[width=\textwidth]{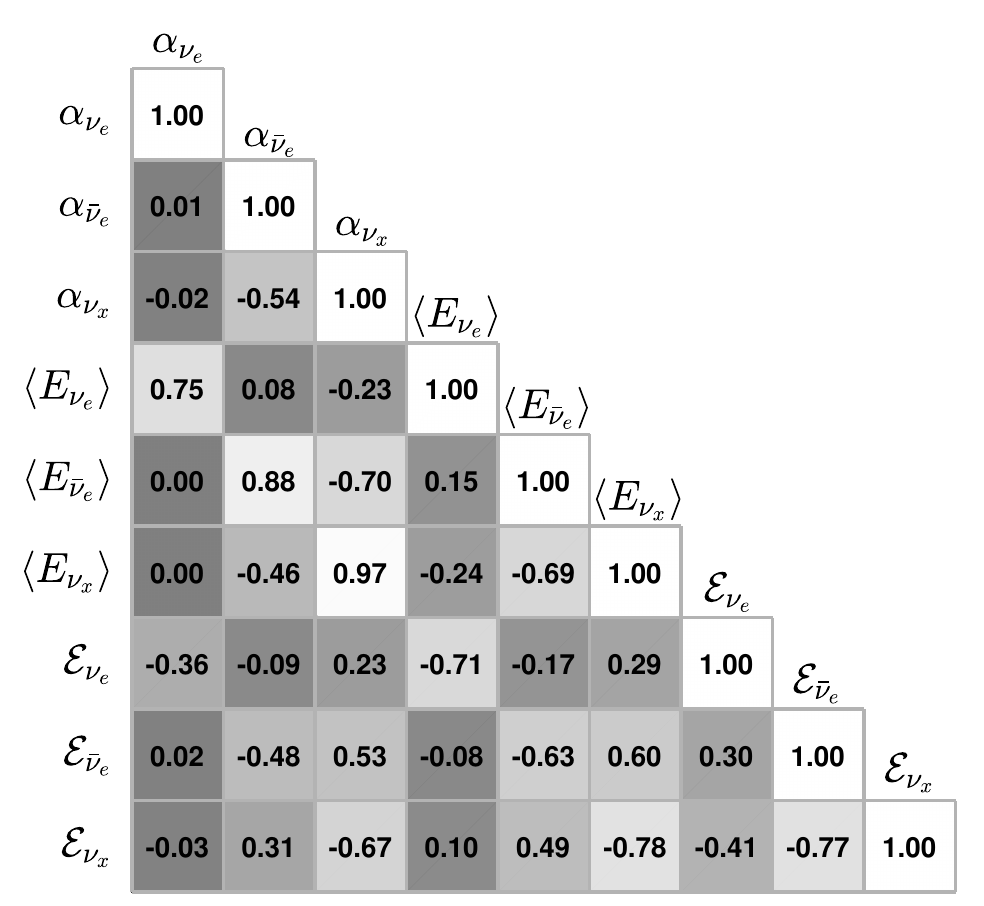}
         \caption{$\mathbf{V}^{\mathrm{NMO}}$.}
         \label{subfig:CorrelationNH}
     \end{subfigure}
     \hfill
     \begin{subfigure}[b]{0.48\textwidth}
         \centering
         \includegraphics[width=\textwidth]{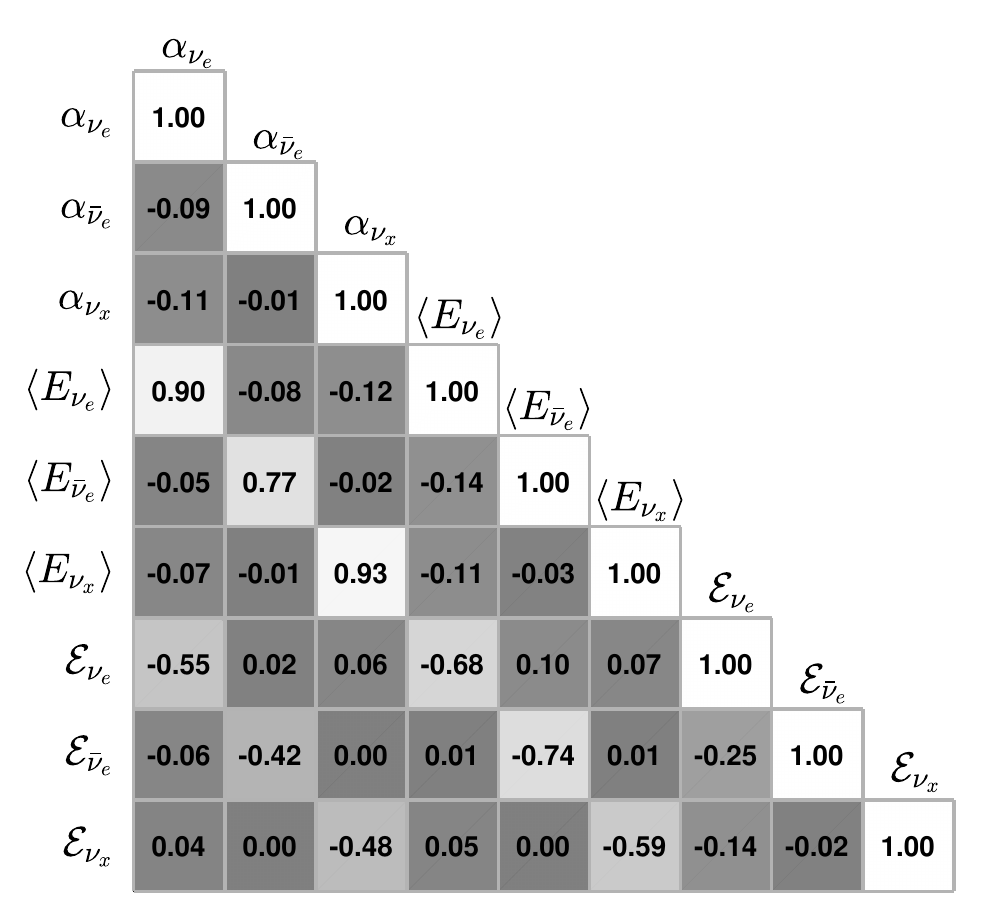}
         \caption{$\mathbf{V}^{\mathrm{IMO}}$.}
         \label{subfig:CorrelationIH}
     \end{subfigure}
        \caption{The matrices of correlation coefficients for NMO and IMO.}
        \label{fig:correlations}
\end{figure}

Aside from the diagonals, the off-diagonal plots in figure~\ref{fig:bayfit_NH_gauss} and figure~\ref{fig:bayfit_IH_gauss} portray the correlations between parameters as 2-dimension distributions.
So as to quantify these correlations, the matrices of correlation coefficients, namely $\mathbf{V}^{\mathrm{NMO}}$ and $\mathbf{V}^{\mathrm{IMO}}$, are calculated and shown in figure~\ref{fig:correlations},
where the value in a coordinate of a matrix refers to the distribution in the same coordinate
of posterior charts.
Apparently, the correlations among the three parameters of one specific species remain the same and, more specifically, another universal hierarchy among the three correlation coefficients emerges as $|\rho(\alpha,\left<E\right>)|>|\rho(\left<E\right>,\mathcal{E})|>|\rho(\alpha,\mathcal{E})|$.
Such patterns are still controlled by the spectral formalism.
On the other hand, different correlation patterns appear between different oscillation models.
In the case of NMO, moderate correlations exist among spectral parameters from $\bar{\nu}_e$ and $\nu_x$ components.
That is, the spectral shape parameters, $\alpha$ and $\left<E\right>$, of $\bar{\nu}_e$ flux have negative correlations to the corresponding parameters of $\nu_x$ flux, and so do the total energy parameters, $\mathcal{E}$.
This can be expected from the mixing of these two components, as described in eq.~\eqref{eq:NH_vebar}.
As a consequence, more complicated correlation patterns stem from two categories of correlations mentioned above (see figure~\ref{subfig:CorrelationNH} and figure~\ref{fig:bayfit_NH_gauss} for more details).
However, it turns out that no such correlations are seen in the case of IMO, while the mixing of $\nu_e$ and $\nu_x$ components does exist, i.e., in eq.~\eqref{eq:IH_ve}.
The absence here is ascribed to the different sensitivities to $\nu_e$ and $\bar{\nu}_e$ species in our detector configurations~\footnote{As a test, the exchange of parameters between $\nu_e$ flavour and $\bar{\nu}_e$ component is estimated again and the mixing-induced correlations are still missing for IMO while clear for NMO. The effect is that these correlations become relatively weaker in the NMO mode. We also swap the values of $\sin^2\theta_{12}$ and $\cos^2\theta_{12}$, and only see some mild effects on the correlation coefficients (even weaker than the previous case).}.
Such difference between NMO and IMO can potentially act as another smoking gun to determine the mass ordering in measurement of the next galactic CCSN~\footnote{When analysing the data with NMO template, dataset with IMO will show even stronger mixing-induced correlations than dataset with NMO (e.g., the correlation coefficients between $\alpha_{\bar{\nu}_e}$ and $\left<E_{\nu_x}\right>$ ($\alpha_{\bar{\nu}_e}$ and $\mathrm{E}_{\nu_x}$) in the two cases are shown as $-0.78$ vs $-0.46$ ($0.64$ vs $0.30$), and, however, the impacts on different coefficients can be different.). If the analyses were conducted with IMO/NO template, we see no manifest signals or just rather weak trends.}, although we postpone further estimates in future work.

Again, we check the results for $d = 5 \kpc$. 
The correlation patterns for both NMO and IMO are still robust, only with modest enhancements found in spectral-induced correlation coefficients of the $\nu_e$ ($\bar{\nu}_e$) flavour in NMO (IMO) conversion.
As to the accuracies of reconstructed parameters, universal improvements of $40\% \sim 50\%$ are again obtained for the $\bar{\nu}_e$ and $\nu_x$ components in the case of NMO, and for the $\nu_e$ and $\nu_x$ components in the case of IMO.
Nevertheless, different parameters of the $\nu_e$ component in NMO conversion show different sensitivities to the change of target distance.
That is, the accuracy for $\mathcal{E}_{\nu_e}$ is increased by $\sim 45\%$ in such test, while that for $\left<E_{\nu_e}\right>$ is only enhanced by $\sim 15\%$ and it turns out to be rather weak improvement ($\sim 4\%$) on $\alpha_{\nu_e}$.
It is similar for that of the $\bar{\nu}_e$ flavour in IMO conversion.
So the measurement of $\alpha_{\nu_e}$ ($\alpha_{\bar{\nu}_e}$) in NMO (IMO) conversion deserves further investigation.

\subsubsection{Impact of cross section uncertainty}

Until now, all the evaluations are performed under an assumption that the cross section for each reaction channel is well determined. 
However, the fact is not so optimistic.
Especially, the cross section for vAr(CC) in DUNE still have large theoretical uncertainty, e.g., a deviation of almost one order of magnitude between the predictions from QRPA-C calculation~\cite{Cheoun:2011zza} and NSM+RPA calculation~\cite{Suzuki:2012ds}.
Such large theoretical uncertainties will substantially impede the interpretation of such measurements, as reported recently by the DUNE Collaboration~\cite{DUNE:2023rtr}.
Here, to draw an idea about how this cross section uncertainty would affect our results, we reanalyse the mock data after introducing a scaling factor of $\pm 5 \%$ on the previous vAr(CC) cross section~\footnote{According to the discussion in Ref.~\cite{DUNE:2023rtr}, we could expect a statistical uncertainty of $5 - 10 \%$ on the total cross section with a ton-scale detector running beside the Spallation Neutron Source for a few years.}. 
The estimated symmetrized fractional uncertainties are also presented in table~\ref{tab:bayfit_results}, just lying below the previous results.
Overall, the total emitted energy is the most sensitive one among the three parameters.
The effects on the other spectral shape parameters are relatively small.
On the other hand, the extraction of neutrino species, which is associated with the data from Hyper-K, will be less influenced by such a cross section uncertainty, since the data from Hyper-K is much more informative than the others.
In particular, the results are almost unaffected in our estimate for $\Bar{\nu}_e$ flux under no oscillation and $\nu_x$ flux under IMO conversion.
Interested readers can refer to table~\ref{tab:bayfit_results} for detailed information.
In all, it can be seen that even a $\pm 5 \%$ uncertainty on the vAr(CC) cross section would manifestly affect the extraction of neutrino spectral parameters here.
And the almost $100 \%$ cross section uncertainty, currently existed, would cause severe biases which make the results not reliable.
So it would be invaluable to control the theoretical precision via a direct measurement of this reaction in the future.

\section{Conclusions}
\label{sec:Con}

In this paper, we present the retrieval of energy spectra for all flavours supernova neutrinos with Bayesian inference by combining data from multiple detectors.
When selecting reaction channels, the collection of IBD-p and eES reactions in Hyper-K, vAr(CC) in DUNE and vPb(NC) in RES-NOVA is employed under the consideration of flavour sensitivity and data statistics.
Before analysing the mock data, we quantify the prior knowledge on the energy spectra of supernova neutrinos with modified Gaussian functions.
Then, using a Poisson likelihood,  
we sample the posterior distribution, which has 9 degrees of freedom, and extract the probability distribution of each parameter. Furthermore, the correlation coefficients among parameters are also estimated and discussed.

Assuming a typical source distance (i.e. $d = 10 \kpc$) in our Galaxy, our results show that the average energy and individual emitted energy can be determined with an accuracy of a few percent in normal (inverted) mass ordering, except for the $\nu_e$ ($\bar{\nu}_e$). Especially, those for heavy flavour neutrinos are reconstructed with a $1 \%$ precision under the oscillation effect of inverted mass ordering. The spectral pinching for either $\bar{\nu}_e$ ($\nu_x$) can also be measured to a few percent precision in normal (inverted) mass ordering.
In contrast, that for either $\nu_e$ or $\bar{\nu}_e$ is hardly extractable from the data, accordingly.
Nevertheless, based on the overall accuracy inferred here, it is interesting to mention that the precise determination of neutron skin of Lead should be promising through nearby galactic supernova neutrino detection in RES-NOVA as proposed in our previous work~\cite{Huang:2022wqu}.
Furthermore, our analyses indicate that there exist two categories of correlations among parameters: spectral-induced correlation and mixing-induced correlation.
The former is encoded in the formalism of neutrino flux, while the latter derives from the complementary effects of neutrino mixing and detector configurations.
Such correlations potentially offer us new ways to extract information from data, more efficiently, via specific combinations of spectral parameters.
It is also possible to solve the mass ordering problem by analysing the mixing-induced correlations.
However, more realistic oscillation models should be included in real observations, e.g., non-adiabatic oscillation, collective oscillation and Earth matter effect.

For future studies, an effective way to enhance the capability of our method is to further improve the flavour-blind sensitivity in the collections (e.g. higher statistics or extra sensitivity to neutrinos with energy below 10 MeV). 
For instance, the neutral current scatterings on $^{16}\mathrm{O}$ in Hyper-K can provide valuable information in the low energy region (i.e., $5 \sim 10 \MeV$), while the pES reaction in JUNO and neutral current scattering on Ar ($\nu + \rm{Ar} \to \nu + \rm{Ar}^\ast$) in DUNE (if available) will offer extra events in the relatively higher energy range. 
It is also worthy to mention that the next-generation large-scale dark matter detectors will also render complementary information in such studies (see, e.g., Ref~\cite{Lang:2016zhv,DarkSide20k:2020ymr}).
Nevertheless, both cross section and detector uncertainties should get treated properly in analysing the data from future realistic observations.
Especially, the cross section uncertainties of vAr(CC) can manifestly affect the results, as shown in our estimate (also see Ref~\cite{DUNE:2023rtr}).
The lack of proper treatment on such uncertainty information makes the spectral parameter uncertainties in the given results, to some extent,  not realistic, which appears as one of the main limitations in this study.
At last, we would like to note that a precise measurement of the distance to supernova is also of great importance, since that would lead to uncertainties on the determination of $\mathcal{E}_\nu$ (see eq.~\eqref{eq:vFluence}).


\appendix

\section{Flat prior}
\label{apd:priors}

\begin{figure}[htbp]
	\includegraphics[width=\textwidth]{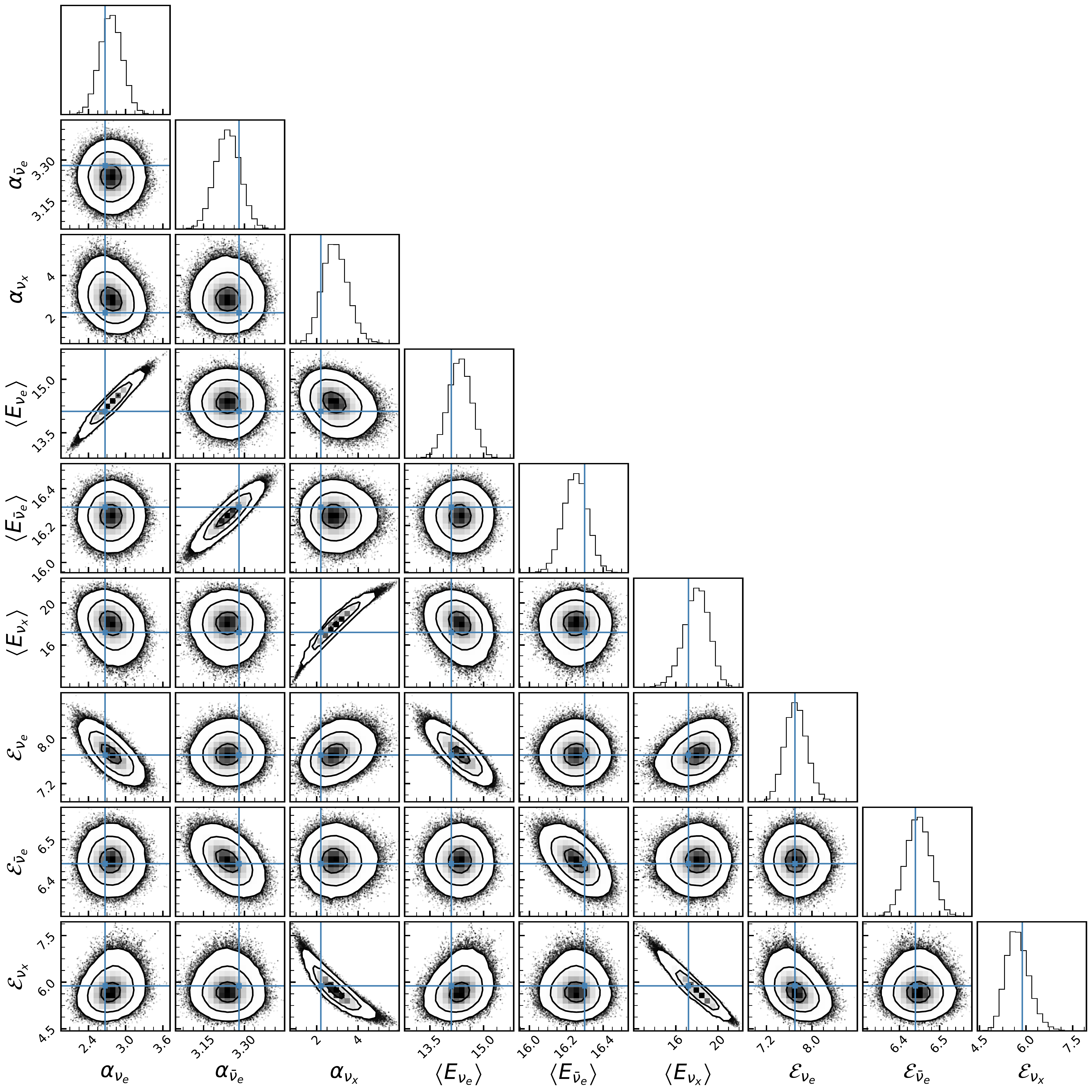}
    \caption{The same as figure~\ref{fig:bayfit_NO_gauss}, but the flat prior distributions are involved.}
    \label{fig:bayfit_NO_flat}
\end{figure}

We replace the Gauss distributions (see, e.g., eq.~\eqref{eq:prior_gauss}) with flat distributions, whose parameter spaces are restricted to the $3 \sigma$ regions of Gauss distributions, in the analysis.
Considering no neutrino oscillation, the results are presented in table~\ref{tab:bayfit_flat} and figure~\ref{fig:bayfit_NO_flat}.
\begin{table}[htbp]
  \caption{The same as table~\ref{tab:bayfit_results}, but flat priors are under estimate now. No neutrino oscillation is assumed.}
  \label{tab:bayfit_flat}
  \centering
  \begin{threeparttable}[b]
     \begin{tabular}{ccccccccccc}
      \toprule
      \multirow{2}*{Osc} & \multirow{2}*{estimate} & \multicolumn{3}{c}{$\alpha$} & \multicolumn{3}{c}{$\left< E \right>$ [MeV]} & \multicolumn{3}{c}{$\mathcal{E}$ [$10^{52} \erg$]} \\
      \cmidrule(lr){3-5}\cmidrule(lr){6-8}\cmidrule(lr){9-11}
      & & $\nu_e$ & $\bar{\nu}_e$ & $\nu_x$ & $\nu_e$ & $\bar{\nu}_e$ & $\nu_x$ & $\nu_e$ & $\bar{\nu}_e$ & $\nu_x$  \\ 
      \midrule
      \multirow{4}{*}{NO} & MAP
      & 2.79 & 3.24 & 2.83 & 14.39 &  16.25 & 17.89 & 7.70 & 6.45 & 5.69 \\
      & $2 \sigma^-$ &  -0.39 & -0.09 & -1.16 & -0.73 & -0.13 & -2.22 & -0.37 & -0.06 & -0.66 \\
      & $2 \sigma^+$ & +0.33 & +0.09 & +1.30 & +0.57 & +0.13 & +2.54 & +0.40 & +0.06 & +0.84 \\
      & $\%$ & 12.9 & 2.8 & 43.5 & 4.5 & 0.8 & 13.3 & 5.0 & 0.9 & 13.2 \\
      \bottomrule
    \end{tabular}
  \end{threeparttable}
\end{table}
Generally speaking, the posterior distributions are quite similar to that of Gauss priors (see, e.g., table~\ref{tab:bayfit_results} and figure~\ref{fig:bayfit_NO_gauss}).
The results of $\bar{\nu}_e$ flavour remain almost the same, due to the highly informative dataset offered by the IBD-p reaction in Hyper-K. 
The influence on the extraction of $\nu_e$ part are also tiny, i.e., only an increase of $\sim 0.3 \%$ on the $2 \sigma$ symmetrized fractional uncertainty. 
However, such replacement shows relatively noticeable impact on the retrieval of $\nu_x$ component, namely an increase of $10.1 \%$ on $\alpha$ and enlargement of $\sim 2.5 \%$ on $\left<E\right>$ and $\mathcal{E}$.
Such consequences are totally reasonable, and confirm the previous statement that the more informative the dataset is, the less dependence the posterior will show on the prior.
Note that these priors can be further updated according to future developments on modelling of stellar core collapse.


\acknowledgments
We are grateful to Ming-chung Chu for useful comments.
X.-R. Huang acknowledges support from Shanghai Jiao Tong University via the Fellowship of Outstanding PhD Graduates.
This work was supported in part by the National Natural Science Foundation of China under Grant Nos. 12235010 and 11625521, and the National SKA Program of China No. 2020SKA0120300.

\paragraph{Note added.} The data and code underlying this article will be shared on reasonable request.


 \bibliographystyle{JHEP}
 \bibliography{biblio.bib}

\providecommand{\href}[2]{#2}\begingroup\raggedright\begin{thebibliography}{10}

\bibitem{Kamiokande-II:1987idp}
{\scshape Kamiokande-II} collaboration, \emph{{Observation of a Neutrino Burst
  from the Supernova SN 1987a}},
  \href{https://doi.org/10.1103/PhysRevLett.58.1490}{\emph{Phys. Rev. Lett.}
  {\bfseries 58} (1987) 1490}.

\bibitem{Bionta:1987qt}
R.M.~Bionta et~al., \emph{{Observation of a Neutrino Burst in Coincidence with
  Supernova SN 1987a in the Large Magellanic Cloud}},
  \href{https://doi.org/10.1103/PhysRevLett.58.1494}{\emph{Phys. Rev. Lett.}
  {\bfseries 58} (1987) 1494}.

\bibitem{ALEXEYEV1988209}
E.~Alexeyev, L.~Alexeyeva, I.~Krivosheina and V.~Volchenko, \emph{Detection of
  the neutrino signal from sn 1987a in the lmc using the inr baksan underground
  scintillation telescope},
  \href{https://doi.org/https://doi.org/10.1016/0370-2693(88)91651-6}{\emph{Physics
  Letters B} {\bfseries 205} (1988) 209}.

\bibitem{Sato:1987rd}
K.~Sato and H.~Suzuki, \emph{{Analysis of Neutrino Burst From the Supernova in
  {LMC}}}, \href{https://doi.org/10.1103/PhysRevLett.58.2722}{\emph{Phys. Rev.
  Lett.} {\bfseries 58} (1987) 2722}.

\bibitem{Burrows:1987zz}
A.~Burrows and J.M.~Lattimer, \emph{{Neutrinos from SN 1987A}},
  \href{https://doi.org/10.1086/184938}{\emph{Astrophys. J. Lett.} {\bfseries
  318} (1987) L63}.

\bibitem{Arnett:1987iz}
W.D.~Arnett and J.L.~Rosner, \emph{{Neutrino Mass Limits From Sn1987a}},
  \href{https://doi.org/10.1103/PhysRevLett.58.1906}{\emph{Phys. Rev. Lett.}
  {\bfseries 58} (1987) 1906}.

\bibitem{Bahcall:1987ua}
J.N.~Bahcall, T.~Piran, W.H.~Press and D.N.~Spergel, \emph{{Neutrino
  Temperatures and Fluxes From the {LMC} Supernova}},
  \href{https://doi.org/10.1038/327682a0}{\emph{Nature} {\bfseries 327} (1987)
  682}.

\bibitem{Loredo:1988mk}
T.J.~Loredo and D.Q.~Lamb, \emph{{Neutrino from SN1987A: Implications for
  cooling of the nascent neutron star and the mass of the electron
  anti-neutrino}},
  \href{https://doi.org/10.1111/j.1749-6632.1989.tb50547.x}{\emph{Annals N. Y.
  Acad. Sci.} {\bfseries 571} (1989) 601}.

\bibitem{Loredo:2001rx}
T.J.~Loredo and D.Q.~Lamb, \emph{{Bayesian analysis of neutrinos observed from
  supernova SN-1987A}},
  \href{https://doi.org/10.1103/PhysRevD.65.063002}{\emph{Phys. Rev. D}
  {\bfseries 65} (2002) 063002}
  [\href{https://arxiv.org/abs/astro-ph/0107260}{{\ttfamily
  astro-ph/0107260}}].

\bibitem{Bethe:1990mw}
H.A.~Bethe, \emph{{Supernova mechanisms}},
  \href{https://doi.org/10.1103/RevModPhys.62.801}{\emph{Rev. Mod. Phys.}
  {\bfseries 62} (1990) 801}.

\bibitem{Janka:2012wk}
H.-T.~Janka, \emph{{Explosion Mechanisms of Core-Collapse Supernovae}},
  \href{https://doi.org/10.1146/annurev-nucl-102711-094901}{\emph{Ann. Rev.
  Nucl. Part. Sci.} {\bfseries 62} (2012) 407}
  [\href{https://arxiv.org/abs/1206.2503}{{\ttfamily 1206.2503}}].

\bibitem{Muller:2016izw}
B.~M\"uller, \emph{{The Status of Multi-Dimensional Core-Collapse Supernova
  Models}}, \href{https://doi.org/10.1017/pasa.2016.40}{\emph{Publ. Astron.
  Soc. Austral.} {\bfseries 33} (2016) e048}
  [\href{https://arxiv.org/abs/1608.03274}{{\ttfamily 1608.03274}}].

\bibitem{OConnor:2018sti}
E.~O'Connor et~al., \emph{{Global Comparison of Core-Collapse Supernova
  Simulations in Spherical Symmetry}},
  \href{https://doi.org/10.1088/1361-6471/aadeae}{\emph{J. Phys. G} {\bfseries
  45} (2018) 104001} [\href{https://arxiv.org/abs/1806.04175}{{\ttfamily
  1806.04175}}].

\bibitem{Just:2018djz}
O.~Just, R.~Bollig, H.-T.~Janka, M.~Obergaulinger, R.~Glas and S.~Nagataki,
  \emph{{Core-collapse supernova simulations in one and two dimensions:
  comparison of codes and approximations}},
  \href{https://doi.org/10.1093/mnras/sty2578}{\emph{Mon. Not. Roy. Astron.
  Soc.} {\bfseries 481} (2018) 4786}
  [\href{https://arxiv.org/abs/1805.03953}{{\ttfamily 1805.03953}}].

\bibitem{Burrows:2020qrp}
A.~Burrows and D.~Vartanyan, \emph{{Core-Collapse Supernova Explosion Theory}},
  \href{https://doi.org/10.1038/s41586-020-03059-w}{\emph{Nature} {\bfseries
  589} (2021) 29} [\href{https://arxiv.org/abs/2009.14157}{{\ttfamily
  2009.14157}}].

\bibitem{Xing:2020ijf}
Z.-z.~Xing, \emph{{Flavor structures of charged fermions and massive
  neutrinos}}, \href{https://doi.org/10.1016/j.physrep.2020.02.001}{\emph{Phys.
  Rept.} {\bfseries 854} (2020) 1}
  [\href{https://arxiv.org/abs/1909.09610}{{\ttfamily 1909.09610}}].

\bibitem{Scholberg:2012id}
K.~Scholberg, \emph{{Supernova Neutrino Detection}},
  \href{https://doi.org/10.1146/annurev-nucl-102711-095006}{\emph{Ann. Rev.
  Nucl. Part. Sci.} {\bfseries 62} (2012) 81}
  [\href{https://arxiv.org/abs/1205.6003}{{\ttfamily 1205.6003}}].

\bibitem{Mosel:2016cwa}
U.~Mosel, \emph{{Neutrino Interactions with Nucleons and Nuclei: Importance for
  Long-Baseline Experiments}},
  \href{https://doi.org/10.1146/annurev-nucl-102115-044720}{\emph{Ann. Rev.
  Nucl. Part. Sci.} {\bfseries 66} (2016) 171}
  [\href{https://arxiv.org/abs/1602.00696}{{\ttfamily 1602.00696}}].

\bibitem{Dutta:2019oaj}
B.~Dutta and L.E.~Strigari, \emph{{Neutrino physics with dark matter
  detectors}},
  \href{https://doi.org/10.1146/annurev-nucl-101918-023450}{\emph{Ann. Rev.
  Nucl. Part. Sci.} {\bfseries 69} (2019) 137}
  [\href{https://arxiv.org/abs/1901.08876}{{\ttfamily 1901.08876}}].

\bibitem{Mirizzi:2015eza}
A.~Mirizzi, I.~Tamborra, H.-T.~Janka, N.~Saviano, K.~Scholberg, R.~Bollig
  et~al., \emph{{Supernova Neutrinos: Production, Oscillations and Detection}},
  \href{https://doi.org/10.1393/ncr/i2016-10120-8}{\emph{Riv. Nuovo Cim.}
  {\bfseries 39} (2016) 1} [\href{https://arxiv.org/abs/1508.00785}{{\ttfamily
  1508.00785}}].

\bibitem{Horiuchi:2018ofe}
S.~Horiuchi and J.P.~Kneller, \emph{{What can be learned from a future
  supernova neutrino detection?}},
  \href{https://doi.org/10.1088/1361-6471/aaa90a}{\emph{J. Phys. G} {\bfseries
  45} (2018) 043002} [\href{https://arxiv.org/abs/1709.01515}{{\ttfamily
  1709.01515}}].

\bibitem{Muller:2019upo}
B.~M\"uller, \emph{{Neutrino Emission as Diagnostics of Core-Collapse
  Supernovae}},
  \href{https://doi.org/10.1146/annurev-nucl-101918-023434}{\emph{Ann. Rev.
  Nucl. Part. Sci.} {\bfseries 69} (2019) 253}
  [\href{https://arxiv.org/abs/1904.11067}{{\ttfamily 1904.11067}}].

\bibitem{Huang:2022wqu}
X.-R.~Huang and L.-W.~Chen, \emph{{Supernova neutrinos as a precise probe of
  nuclear neutron skin}},
  \href{https://doi.org/10.1103/PhysRevD.106.123034}{\emph{Phys. Rev. D}
  {\bfseries 106} (2022) 123034}
  [\href{https://arxiv.org/abs/2210.04534}{{\ttfamily 2210.04534}}].

\bibitem{Chauhan:2022wgj}
B.~Chauhan, \emph{{Using supernova neutrinos to probe strange spin of proton
  with JUNO and THEIA}},  \href{https://arxiv.org/abs/2211.08443}{{\ttfamily
  2211.08443}}.

\bibitem{Baum:2022wfc}
S.~Baum, F.~Capozzi and S.~Horiuchi, \emph{{Rocks, water, and noble liquids:
  Unfolding the flavor contents of supernova neutrinos}},
  \href{https://doi.org/10.1103/PhysRevD.106.123008}{\emph{Phys. Rev. D}
  {\bfseries 106} (2022) 123008}
  [\href{https://arxiv.org/abs/2203.12696}{{\ttfamily 2203.12696}}].

\bibitem{OConnor:2018tuw}
E.P.~O'Connor and S.M.~Couch, \emph{{Exploring Fundamentally Three-dimensional
  Phenomena in High-fidelity Simulations of Core-collapse Supernovae}},
  \href{https://doi.org/10.3847/1538-4357/aadcf7}{\emph{Astrophys. J.}
  {\bfseries 865} (2018) 81}
  [\href{https://arxiv.org/abs/1807.07579}{{\ttfamily 1807.07579}}].

\bibitem{Burrows:2019zce}
A.~Burrows, D.~Radice, D.~Vartanyan, H.~Nagakura, M.A.~Skinner and J.~Dolence,
  \emph{{The Overarching Framework of Core-Collapse Supernova Explosions as
  Revealed by 3D Fornax Simulations}},
  \href{https://doi.org/10.1093/mnras/stz3223}{\emph{Mon. Not. Roy. Astron.
  Soc.} {\bfseries 491} (2020) 2715}
  [\href{https://arxiv.org/abs/1909.04152}{{\ttfamily 1909.04152}}].

\bibitem{Nagakura:2020qhb}
H.~Nagakura, A.~Burrows, D.~Vartanyan and D.~Radice, \emph{{Core-collapse
  supernova neutrino emission and detection informed by state-of-the-art
  three-dimensional numerical models}},
  \href{https://doi.org/10.1093/mnras/staa2691}{\emph{Mon. Not. Roy. Astron.
  Soc.} {\bfseries 500} (2020) 696}
  [\href{https://arxiv.org/abs/2007.05000}{{\ttfamily 2007.05000}}].

\bibitem{Hyper-Kamiokande:2018ofw}
{\scshape Hyper-Kamiokande} collaboration, \emph{{Hyper-Kamiokande Design
  Report}},  \href{https://arxiv.org/abs/1805.04163}{{\ttfamily 1805.04163}}.

\bibitem{IceCube:2011cwc}
{\scshape IceCube} collaboration, \emph{{IceCube Sensitivity for Low-Energy
  Neutrinos from Nearby Supernovae}},
  \href{https://doi.org/10.1051/0004-6361/201117810e}{\emph{Astron. Astrophys.}
  {\bfseries 535} (2011) A109}
  [\href{https://arxiv.org/abs/1108.0171}{{\ttfamily 1108.0171}}].

\bibitem{JUNO:2015zny}
{\scshape JUNO} collaboration, \emph{{Neutrino Physics with JUNO}},
  \href{https://doi.org/10.1088/0954-3899/43/3/030401}{\emph{J. Phys. G}
  {\bfseries 43} (2016) 030401}
  [\href{https://arxiv.org/abs/1507.05613}{{\ttfamily 1507.05613}}].

\bibitem{Theia:2019non}
{\scshape Theia} collaboration, \emph{{THEIA: an advanced optical neutrino
  detector}}, \href{https://doi.org/10.1140/epjc/s10052-020-7977-8}{\emph{Eur.
  Phys. J. C} {\bfseries 80} (2020) 416}
  [\href{https://arxiv.org/abs/1911.03501}{{\ttfamily 1911.03501}}].

\bibitem{DUNE:2020lwj}
{\scshape DUNE} collaboration, \emph{{Deep Underground Neutrino Experiment
  (DUNE), Far Detector Technical Design Report, Volume I Introduction to
  DUNE}}, \href{https://doi.org/10.1088/1748-0221/15/08/T08008}{\emph{JINST}
  {\bfseries 15} (2020) T08008}
  [\href{https://arxiv.org/abs/2002.02967}{{\ttfamily 2002.02967}}].

\bibitem{DUNE:2020ypp}
{\scshape DUNE} collaboration, \emph{{Deep Underground Neutrino Experiment
  (DUNE), Far Detector Technical Design Report, Volume II: DUNE Physics}},
  \href{https://arxiv.org/abs/2002.03005}{{\ttfamily 2002.03005}}.

\bibitem{DUNE:2020zfm}
{\scshape DUNE} collaboration, \emph{{Supernova neutrino burst detection with
  the Deep Underground Neutrino Experiment}},
  \href{https://doi.org/10.1140/epjc/s10052-021-09166-w}{\emph{Eur. Phys. J. C}
  {\bfseries 81} (2021) 423}
  [\href{https://arxiv.org/abs/2008.06647}{{\ttfamily 2008.06647}}].

\bibitem{Pattavina:2020cqc}
L.~Pattavina, N.~Ferreiro~Iachellini and I.~Tamborra, \emph{{Neutrino
  observatory based on archaeological lead}},
  \href{https://doi.org/10.1103/PhysRevD.102.063001}{\emph{Phys. Rev. D}
  {\bfseries 102} (2020) 063001}
  [\href{https://arxiv.org/abs/2004.06936}{{\ttfamily 2004.06936}}].

\bibitem{RES-NOVA:2021gqp}
{\scshape RES-NOVA} collaboration, \emph{{RES-NOVA sensitivity to core-collapse
  and failed core-collapse supernova neutrinos}},
  \href{https://doi.org/10.1088/1475-7516/2021/10/064}{\emph{JCAP} {\bfseries
  10} (2021) 064} [\href{https://arxiv.org/abs/2103.08672}{{\ttfamily
  2103.08672}}].

\bibitem{Rozwadowska:2020nab}
K.~Rozwadowska, F.~Vissani and E.~Cappellaro, \emph{{On the rate of core
  collapse supernovae in the milky way}},
  \href{https://doi.org/10.1016/j.newast.2020.101498}{\emph{New Astron.}
  {\bfseries 83} (2021) 101498}
  [\href{https://arxiv.org/abs/2009.03438}{{\ttfamily 2009.03438}}].

\bibitem{Laha:2013hva}
R.~Laha and J.F.~Beacom, \emph{{Gadolinium in water Cherenkov detectors
  improves detection of supernova $\nu_e$}},
  \href{https://doi.org/10.1103/PhysRevD.89.063007}{\emph{Phys. Rev. D}
  {\bfseries 89} (2014) 063007}
  [\href{https://arxiv.org/abs/1311.6407}{{\ttfamily 1311.6407}}].

\bibitem{Nikrant:2017nya}
A.~Nikrant, R.~Laha and S.~Horiuchi, \emph{{Robust measurement of supernova
  $\nu_e$ spectra with future neutrino detectors}},
  \href{https://doi.org/10.1103/PhysRevD.97.023019}{\emph{Phys. Rev. D}
  {\bfseries 97} (2018) 023019}
  [\href{https://arxiv.org/abs/1711.00008}{{\ttfamily 1711.00008}}].

\bibitem{Dasgupta:2011wg}
B.~Dasgupta and J.F.~Beacom, \emph{{Reconstruction of supernova $\nu_\mu$,
  $\nu_\tau$, anti-$\nu_\mu$, and anti-$\nu_\tau$ neutrino spectra at
  scintillator detectors}},
  \href{https://doi.org/10.1103/PhysRevD.83.113006}{\emph{Phys. Rev. D}
  {\bfseries 83} (2011) 113006}
  [\href{https://arxiv.org/abs/1103.2768}{{\ttfamily 1103.2768}}].

\bibitem{Lu:2016ipr}
J.-S.~Lu, Y.-F.~Li and S.~Zhou, \emph{{Getting the most from the detection of
  Galactic supernova neutrinos in future large liquid-scintillator detectors}},
  \href{https://doi.org/10.1103/PhysRevD.94.023006}{\emph{Phys. Rev. D}
  {\bfseries 94} (2016) 023006}
  [\href{https://arxiv.org/abs/1605.07803}{{\ttfamily 1605.07803}}].

\bibitem{GalloRosso:2017mdz}
A.~Gallo~Rosso, F.~Vissani and M.C.~Volpe, \emph{{What can we learn on
  supernova neutrino spectra with water Cherenkov detectors?}},
  \href{https://doi.org/10.1088/1475-7516/2018/04/040}{\emph{JCAP} {\bfseries
  04} (2018) 040} [\href{https://arxiv.org/abs/1712.05584}{{\ttfamily
  1712.05584}}].

\bibitem{GalloRosso:2020qqa}
A.~Gallo~Rosso, \emph{{Supernova neutrino fluxes in HALO-1kT, Super-Kamiokande,
  and~JUNO}}, \href{https://doi.org/10.1088/1475-7516/2021/06/046}{\emph{JCAP}
  {\bfseries 06} (2021) 046}
  [\href{https://arxiv.org/abs/2012.12579}{{\ttfamily 2012.12579}}].

\bibitem{Li:2017dbg}
H.-L.~Li, Y.-F.~Li, M.~Wang, L.-J.~Wen and S.~Zhou, \emph{{Towards a complete
  reconstruction of supernova neutrino spectra in future large
  liquid-scintillator detectors}},
  \href{https://doi.org/10.1103/PhysRevD.97.063014}{\emph{Phys. Rev. D}
  {\bfseries 97} (2018) 063014}
  [\href{https://arxiv.org/abs/1712.06985}{{\ttfamily 1712.06985}}].

\bibitem{Li:2019qxi}
H.-L.~Li, X.~Huang, Y.-F.~Li, L.-J.~Wen and S.~Zhou, \emph{{Model-independent
  approach to the reconstruction of multiflavor supernova neutrino energy
  spectra}}, \href{https://doi.org/10.1103/PhysRevD.99.123009}{\emph{Phys. Rev.
  D} {\bfseries 99} (2019) 123009}
  [\href{https://arxiv.org/abs/1903.04781}{{\ttfamily 1903.04781}}].

\bibitem{Nagakura:2020bbw}
H.~Nagakura, \emph{{Retrieval of energy spectra for all flavours of neutrinos
  from core-collapse supernova with multiple detectors}},
  \href{https://doi.org/10.1093/mnras/staa3287}{\emph{Mon. Not. Roy. Astron.
  Soc.} {\bfseries 500} (2020) 319}
  [\href{https://arxiv.org/abs/2008.10082}{{\ttfamily 2008.10082}}].

\bibitem{Keil:2002in}
M.T.~Keil, G.G.~Raffelt and H.-T.~Janka, \emph{{Monte Carlo study of supernova
  neutrino spectra formation}},
  \href{https://doi.org/10.1086/375130}{\emph{Astrophys. J.} {\bfseries 590}
  (2003) 971} [\href{https://arxiv.org/abs/astro-ph/0208035}{{\ttfamily
  astro-ph/0208035}}].

\bibitem{Tamborra:2012ac}
I.~Tamborra, B.~Muller, L.~Hudepohl, H.-T.~Janka and G.~Raffelt,
  \emph{{High-resolution supernova neutrino spectra represented by a simple
  fit}}, \href{https://doi.org/10.1103/PhysRevD.86.125031}{\emph{Phys. Rev. D}
  {\bfseries 86} (2012) 125031}
  [\href{https://arxiv.org/abs/1211.3920}{{\ttfamily 1211.3920}}].

\bibitem{DAgostini:2003bpu}
G.~D'Agostini, \emph{{Bayesian inference in processing experimental data:
  Principles and basic applications}},
  \href{https://doi.org/10.1088/0034-4885/66/9/201}{\emph{Rept. Prog. Phys.}
  {\bfseries 66} (2003) 1383}
  [\href{https://arxiv.org/abs/physics/0304102}{{\ttfamily physics/0304102}}].

\bibitem{Ashton:2018jfp}
G.~Ashton et~al., \emph{{BILBY: A user-friendly Bayesian inference library for
  gravitational-wave astronomy}},
  \href{https://doi.org/10.3847/1538-4365/ab06fc}{\emph{Astrophys. J. Suppl.}
  {\bfseries 241} (2019) 27}
  [\href{https://arxiv.org/abs/1811.02042}{{\ttfamily 1811.02042}}].

\bibitem{Bernhard:2016tnd}
J.E.~Bernhard, J.S.~Moreland, S.A.~Bass, J.~Liu and U.~Heinz, \emph{{Applying
  Bayesian parameter estimation to relativistic heavy-ion collisions:
  simultaneous characterization of the initial state and quark-gluon plasma
  medium}}, \href{https://doi.org/10.1103/PhysRevC.94.024907}{\emph{Phys. Rev.
  C} {\bfseries 94} (2016) 024907}
  [\href{https://arxiv.org/abs/1605.03954}{{\ttfamily 1605.03954}}].

\bibitem{Trotta:2008qt}
R.~Trotta, \emph{{Bayes in the sky: Bayesian inference and model selection in
  cosmology}}, \href{https://doi.org/10.1080/00107510802066753}{\emph{Contemp.
  Phys.} {\bfseries 49} (2008) 71}
  [\href{https://arxiv.org/abs/0803.4089}{{\ttfamily 0803.4089}}].

\bibitem{Loredo:2012jm}
T.J.~Loredo, \emph{{Bayesian astrostatistics: a backward look to the future}},
  \href{https://arxiv.org/abs/1208.3036}{{\ttfamily 1208.3036}}.

\bibitem{Scholberg:2017czd}
K.~Scholberg, \emph{{Supernova Signatures of Neutrino Mass Ordering}},
  \href{https://doi.org/10.1088/1361-6471/aa97be}{\emph{J. Phys. G} {\bfseries
  45} (2018) 014002} [\href{https://arxiv.org/abs/1707.06384}{{\ttfamily
  1707.06384}}].

\bibitem{Hyper-Kamiokande:2021frf}
{\scshape Hyper-Kamiokande} collaboration, \emph{{Supernova Model
  Discrimination with Hyper-Kamiokande}},
  \href{https://doi.org/10.3847/1538-4357/abf7c4}{\emph{Astrophys. J.}
  {\bfseries 916} (2021) 15}
  [\href{https://arxiv.org/abs/2101.05269}{{\ttfamily 2101.05269}}].

\bibitem{Huang:2019ene}
X.-R.~Huang and L.-W.~Chen, \emph{{Neutron Skin in CsI and Low-Energy Effective
  Weak Mixing Angle from COHERENT Data}},
  \href{https://doi.org/10.1103/PhysRevD.100.071301}{\emph{Phys. Rev. D}
  {\bfseries 100} (2019) 071301}
  [\href{https://arxiv.org/abs/1902.07625}{{\ttfamily 1902.07625}}].

\bibitem{PREX:2021umo}
{\scshape PREX} collaboration, \emph{{Accurate Determination of the Neutron
  Skin Thickness of $^{208}$Pb through Parity-Violation in Electron
  Scattering}},
  \href{https://doi.org/10.1103/PhysRevLett.126.172502}{\emph{Phys. Rev. Lett.}
  {\bfseries 126} (2021) 172502}
  [\href{https://arxiv.org/abs/2102.10767}{{\ttfamily 2102.10767}}].

\bibitem{COHERENT:2017ipa}
{\scshape COHERENT} collaboration, \emph{{Observation of Coherent Elastic
  Neutrino-Nucleus Scattering}},
  \href{https://doi.org/10.1126/science.aao0990}{\emph{Science} {\bfseries 357}
  (2017) 1123} [\href{https://arxiv.org/abs/1708.01294}{{\ttfamily
  1708.01294}}].

\bibitem{COHERENT:2018imc}
{\scshape COHERENT} collaboration, \emph{{COHERENT Collaboration data release
  from the first observation of coherent elastic neutrino-nucleus scattering}},
   \href{https://arxiv.org/abs/1804.09459}{{\ttfamily 1804.09459}}.

\bibitem{O'Connor_2013}
E.~O'Connor and C.D.~Ott, \emph{The progenitor dependence of the pre-explosion
  neutrino emission in core-collapse supernovae},
  \href{https://doi.org/10.1088/0004-637X/762/2/126}{\emph{Astrophys. J.}
  {\bfseries 762} (2012) 126}.

\bibitem{Seadrow:2018ftp}
S.~Seadrow, A.~Burrows, D.~Vartanyan, D.~Radice and M.A.~Skinner,
  \emph{{Neutrino Signals of Core-Collapse Supernovae in Underground
  Detectors}}, \href{https://doi.org/10.1093/mnras/sty2164}{\emph{Mon. Not.
  Roy. Astron. Soc.} {\bfseries 480} (2018) 4710}
  [\href{https://arxiv.org/abs/1804.00689}{{\ttfamily 1804.00689}}].

\bibitem{Pan:2017tpk}
K.-C.~Pan, M.~Liebend\"orfer, S.M.~Couch and F.-K.~Thielemann, \emph{{Equation
  of State Dependent Dynamics and Multi-messenger Signals from Stellar-mass
  Black Hole Formation}},
  \href{https://doi.org/10.3847/1538-4357/aab71d}{\emph{Astrophys. J.}
  {\bfseries 857} (2018) 13}
  [\href{https://arxiv.org/abs/1710.01690}{{\ttfamily 1710.01690}}].

\bibitem{daSilvaSchneider:2020ddu}
A.~da~Silva~Schneider, E.~O'Connor, E.~Granqvist, A.~Betranhandy and
  S.M.~Couch, \emph{{Equation of State and Progenitor Dependence of
  Stellar-mass Black Hole Formation}},
  \href{https://doi.org/10.3847/1538-4357/ab8308}{\emph{Astrophys. J.}
  {\bfseries 894} (2020) 4} [\href{https://arxiv.org/abs/2001.10434}{{\ttfamily
  2001.10434}}].

\bibitem{Raduta:2021coc}
A.R.~Raduta, F.~Nacu and M.~Oertel, \emph{{Equations of state for hot neutron
  stars}}, \href{https://doi.org/10.1140/epja/s10050-021-00628-z}{\emph{Eur.
  Phys. J. A} {\bfseries 57} (2021) 329}
  [\href{https://arxiv.org/abs/2109.00251}{{\ttfamily 2109.00251}}].

\bibitem{Minakata:2008nc}
H.~Minakata, H.~Nunokawa, R.~Tomas and J.W.F.~Valle, \emph{{Parameter
  Degeneracy in Flavor-Dependent Reconstruction of Supernova Neutrino Fluxes}},
  \href{https://doi.org/10.1088/1475-7516/2008/12/006}{\emph{JCAP} {\bfseries
  12} (2008) 006} [\href{https://arxiv.org/abs/0802.1489}{{\ttfamily
  0802.1489}}].

\bibitem{Liao:2016uis}
W.~Liao, \emph{{Detecting supernovae neutrino with Earth matter effect}},
  \href{https://doi.org/10.1103/PhysRevD.94.113016}{\emph{Phys. Rev. D}
  {\bfseries 94} (2016) 113016}
  [\href{https://arxiv.org/abs/1607.03334}{{\ttfamily 1607.03334}}].

\bibitem{Akhmedov:2002zj}
E.K.~Akhmedov, C.~Lunardini and A.Y.~Smirnov, \emph{{Supernova neutrinos:
  Difference of muon-neutrino - tau-neutrino fluxes and conversion effects}},
  \href{https://doi.org/10.1016/S0550-3213(02)00692-2}{\emph{Nucl. Phys. B}
  {\bfseries 643} (2002) 339}
  [\href{https://arxiv.org/abs/hep-ph/0204091}{{\ttfamily hep-ph/0204091}}].

\bibitem{Balantekin:2007es}
A.B.~Balantekin, J.~Gava and C.~Volpe, \emph{{Possible CP-Violation effects in
  core-collapse Supernovae}},
  \href{https://doi.org/10.1016/j.physletb.2008.03.038}{\emph{Phys. Lett. B}
  {\bfseries 662} (2008) 396}
  [\href{https://arxiv.org/abs/0710.3112}{{\ttfamily 0710.3112}}].

\bibitem{ParticleDataGroup:2022pth}
{\scshape Particle Data Group} collaboration, \emph{{Review of Particle
  Physics}}, \href{https://doi.org/10.1093/ptep/ptac097}{\emph{PTEP} {\bfseries
  2022} (2022) 083C01}.

\bibitem{Hudepohl:2009tyy}
L.~Hudepohl, B.~Muller, H.T.~Janka, A.~Marek and G.G.~Raffelt, \emph{{Neutrino
  Signal of Electron-Capture Supernovae from Core Collapse to Cooling}},
  \href{https://doi.org/10.1103/PhysRevLett.104.251101}{\emph{Phys. Rev. Lett.}
  {\bfseries 104} (2010) 251101}
  [\href{https://arxiv.org/abs/0912.0260}{{\ttfamily 0912.0260}}].

\bibitem{Foreman-Mackey:2012any}
D.~Foreman-Mackey, D.W.~Hogg, D.~Lang and J.~Goodman, \emph{{emcee: The MCMC
  Hammer}}, \href{https://doi.org/10.1086/670067}{\emph{Publ. Astron. Soc.
  Pac.} {\bfseries 125} (2013) 306}
  [\href{https://arxiv.org/abs/1202.3665}{{\ttfamily 1202.3665}}].

\bibitem{Mukhopadhyay:2020ubs}
M.~Mukhopadhyay, C.~Lunardini, F.X.~Timmes and K.~Zuber, \emph{{Presupernova
  neutrinos: directional sensitivity and prospects for progenitor
  identification}},
  \href{https://doi.org/10.3847/1538-4357/ab99a6}{\emph{Astrophys. J.}
  {\bfseries 899} (2020) 153}
  [\href{https://arxiv.org/abs/2004.02045}{{\ttfamily 2004.02045}}].

\bibitem{corner}
D.~Foreman-Mackey, \emph{corner.py: Scatterplot matrices in python},
  \href{https://doi.org/10.21105/joss.00024}{\emph{The Journal of Open Source
  Software} {\bfseries 1} (2016) 24}.

\bibitem{Cheoun:2011zza}
M.-K.~Cheoun, E.~Ha and T.~Kajino, \emph{{Reactions on Ar-40 involving solar
  neutrinos and neutrinos from core-collapsing supernovae}},
  \href{https://doi.org/10.1103/PhysRevC.83.028801}{\emph{Phys. Rev. C}
  {\bfseries 83} (2011) 028801}.

\bibitem{Suzuki:2012ds}
T.~Suzuki and M.~Honma, \emph{{Neutrino Capture Reactions on $^{40}$Ar}},
  \href{https://doi.org/10.1103/PhysRevC.87.014607}{\emph{Phys. Rev. C}
  {\bfseries 87} (2013) 014607}
  [\href{https://arxiv.org/abs/1211.4078}{{\ttfamily 1211.4078}}].

\bibitem{DUNE:2023rtr}
{\scshape DUNE} collaboration, \emph{{Impact of cross-section uncertainties on
  supernova neutrino spectral parameter fitting in the Deep Underground
  Neutrino Experiment}},
  \href{https://doi.org/10.1103/PhysRevD.107.112012}{\emph{Phys. Rev. D}
  {\bfseries 107} (2023) 112012}
  [\href{https://arxiv.org/abs/2303.17007}{{\ttfamily 2303.17007}}].

\bibitem{Lang:2016zhv}
R.F.~Lang, C.~McCabe, S.~Reichard, M.~Selvi and I.~Tamborra, \emph{{Supernova
  neutrino physics with xenon dark matter detectors: A timely perspective}},
  \href{https://doi.org/10.1103/PhysRevD.94.103009}{\emph{Phys. Rev. D}
  {\bfseries 94} (2016) 103009}
  [\href{https://arxiv.org/abs/1606.09243}{{\ttfamily 1606.09243}}].

\bibitem{DarkSide20k:2020ymr}
{\scshape DarkSide 20k} collaboration, \emph{{Sensitivity of future liquid
  argon dark matter search experiments to core-collapse supernova neutrinos}},
  \href{https://doi.org/10.1088/1475-7516/2021/03/043}{\emph{JCAP} {\bfseries
  03} (2021) 043} [\href{https://arxiv.org/abs/2011.07819}{{\ttfamily
  2011.07819}}].

\end{thebibliography}\endgroup

\end{document}